**Dr. Petar Radanliev**
Parks Road,
Oxford OX1 3PJ
United Kingdom
Email: petar.radanliev@cs.ox.ac.uk
BA Hons., MSc., Ph.D. Post-Doctorate

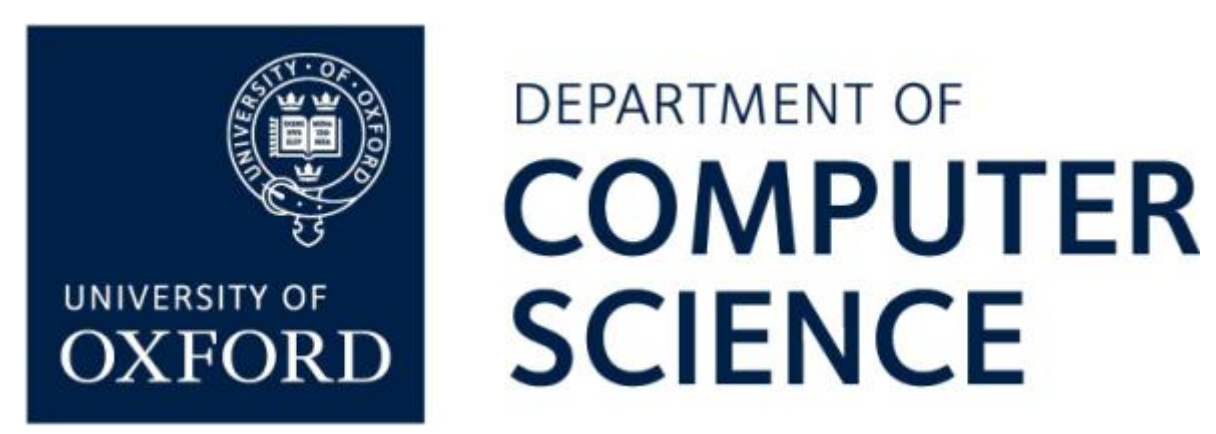

# Execution-Bound Advisory Automation for Agentic AI: A Reproducible AIBOM-Driven CSAF-VEX Framework

**Corresponding author: Petar Radanliev[1]*

Petar Radanliev[1,3], Omar Santos[2], Carsten Maple[3,4], Kay Atefi[5],

[1] Department of Computer Sciences, University of Oxford, Wolfson Building, Parks Rd, Oxford OX1 3QG

[2] Cisco Systems, RTP, North Carolina, United States,

[3] The Alan Turing Institute, British Library, 96 Euston Rd., London NW1 2DB

[4] University of Warwick – WMG, 6 Lord Bhattacharyya Way, Coventry CV4 7AL,

[5] Department of Computer Science, School of Digital & Physical Sciences, Faculty of science and Engineering, University of Hull

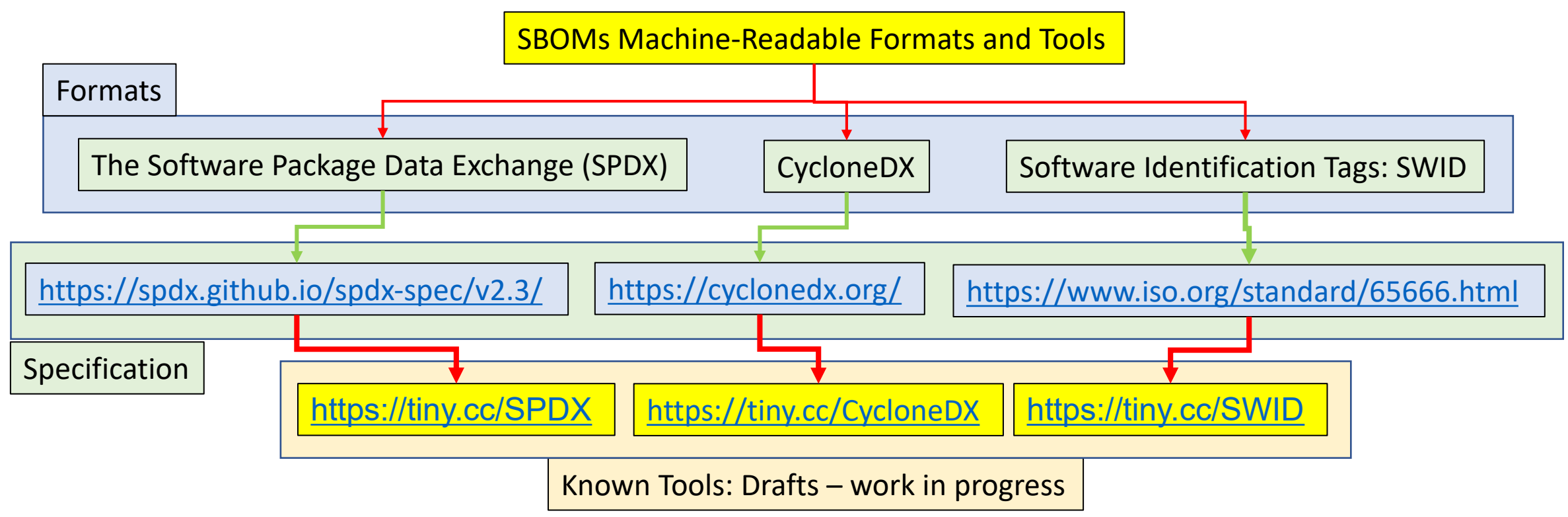

**Dr. Petar Radanliev**
Parks Road,
Oxford OX1 3PJ
United Kingdom
Email: petar.radanliev@cs.ox.ac.uk
BA Hons., MSc., Ph.D. Post-Doctorate

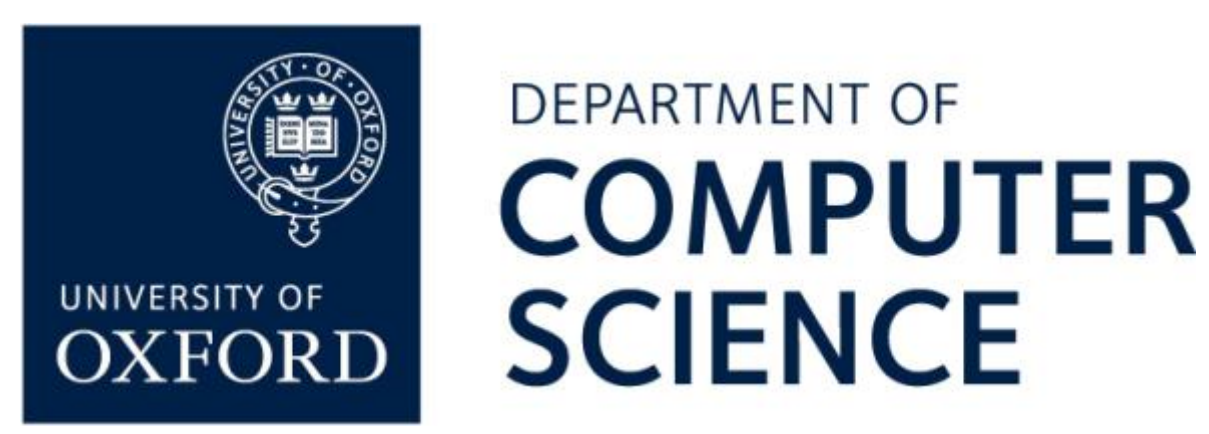

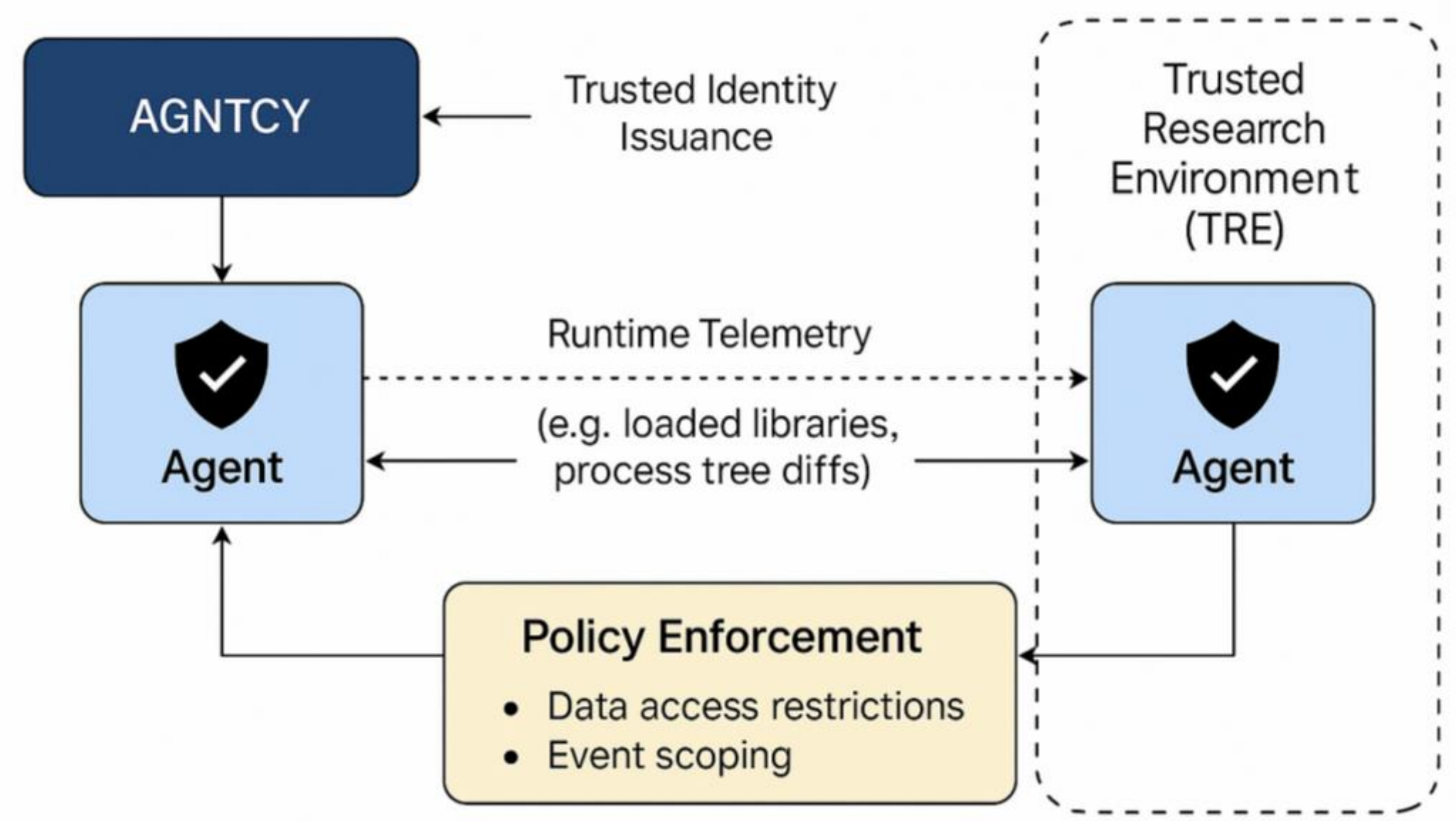

***Abstract:***

**Introduction:** Agentic AI systems integrate foundation models, prompt templates, tool connectors, orchestration logic, and containerised dependencies, creating exploitability conditions that cannot be inferred from static Software Bills of Materials (SBOMs). Artificial Intelligence Bills of Materials (AIBOM) extend transparency to AI-specific artefacts, yet current CSAF/VEX workflows remain based on static component–CVE correlation without runtime validation.

**Materials and Methods:** A protocol-driven framework is presented that binds SBOM and AIBOM artefacts to deterministic environment capture and structured runtime telemetry. Exploitability is computed from declared artefacts, observed activation conditions, and enforced execution policies. CSAF-VEX advisories are generated from combined static and runtime evidence, cryptographically signed, and validated through deterministic replay. Evaluation uses approximately 10,000 component entries across synthetic Agentic AI workloads (50–5,000 components), incorporating OSV, GitHub Advisory, KEV, and EPSS datasets.

**Results:** Under controlled experimental conditions, the framework achieves an F1-score of 0.93 (precision 0.96, recall 0.92), reduces false positives by up to 42% relative to static SBOM–CVE matching without runtime validation, and alters exploitability outcomes in 31% of AI-specific artefact cases through AIBOM extension. Advisory artefacts remain reproducible under deterministic replay.

**Discussion:** Binding AIBOM artefacts to runtime telemetry transforms CSAF-VEX generation from static disclosure into execution-grounded exploitability assessment for Agentic AI supply chains.

**Keywords:** Agentic AI; Artificial Intelligence Bill of Materials (AIBOM); Execution-Bound Exploitability; CSAF-VEX Automation; Runtime Telemetry; AI Supply Chain Security.

**Dr. Petar Radanliev**
Parks Road,
Oxford OX1 3PJ
United Kingdom
Email: petar.radanliev@cs.ox.ac.uk
BA Hons., MSc., Ph.D. Post-Doctorate

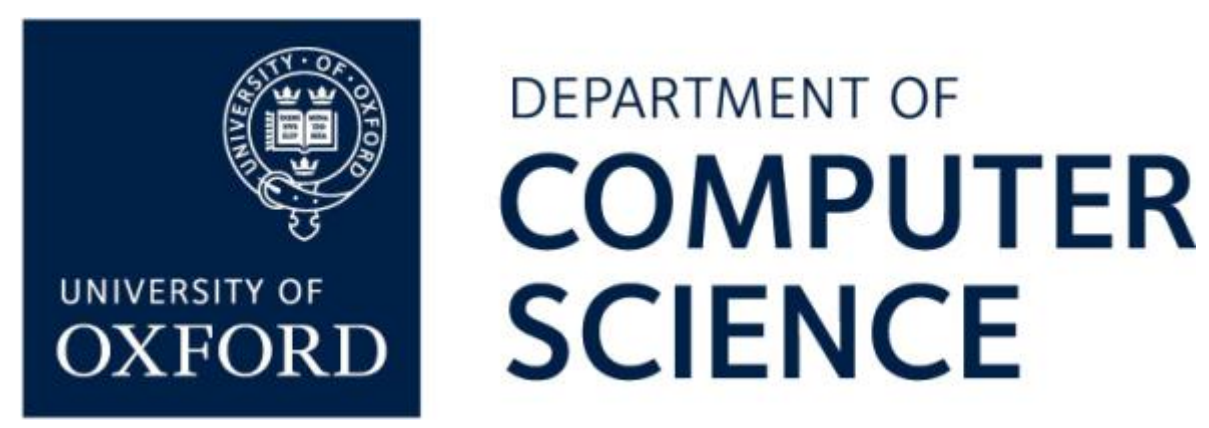

# 1 From Static AIBOM and SBOM Processing to Agentic, Context-Aware Advisory Systems

Software supply chain security has progressed rapidly in recent years, driven by regulatory mandates and the increasing adoption of machine-readable transparency artefacts such as the Software Bill of Materials (SBOM) [1], [2], the Common Security Advisory Framework (CSAF) [3], and the Vulnerability Exploitability eXchange (VEX) [4]. These mechanisms were designed to improve visibility into component dependencies and to reduce the operational burden associated with large-scale vulnerability management. Nevertheless, despite their formal standardisation and growing institutional support, current implementations remain fundamentally static. SBOMs enumerate declared components, CSAF structures advisory distribution, and VEX aims to filter non-exploitable vulnerabilities. Yet these artefacts are typically produced and consumed without verifiable linkage to runtime execution state, operational policy constraints, or the behavioural characteristics of the systems in which components are deployed.

This limitation becomes particularly pronounced in the context of contemporary Agentic AI systems. Modern AI deployments are not monolithic software artefacts but dynamic, tool-using, multi-agent environments composed of foundation models, prompts, orchestration layers, plug-ins, vector stores, external APIs, and runtime dependencies. Their behaviour is shaped not only by static libraries but also by contextual inputs, memory, inter-agent communication, and enforced execution policies. Traditional SBOM-based advisory workflows do not adequately capture model lineage, prompt dependencies, tool invocation chains, policy-bound execution constraints, or agent-to-agent communication surfaces. As a result, exploitability assessments derived from static component inventories frequently overestimate operational risk and generate significant false-positive remediation effort.

Emerging work on Artificial Intelligence Bills of Materials (AIBOM) [5] acknowledges that AI systems require extended transparency models beyond conventional SBOMs. However, while AIBOM initiatives expand the scope of declared artefacts to include models and datasets, they do not yet provide a reproducible and cryptographically verifiable mechanism for binding runtime telemetry, execution context, and advisory generation into a coherent lifecycle. In parallel, CSAF and VEX standards continue to evolve, but their automation pipelines largely rely on static correlation between component identifiers and vulnerability databases. The consequence is a structural gap between enumerated vulnerabilities and context-resolved exploitability.

A central unresolved challenge, therefore, lies in transforming advisory generation from a static enumeration process into a context-aware, agent-mediated, and reproducible computational workflow. Exploitability in operational systems depends not only on the presence of vulnerable components but also on runtime activation conditions, environmental constraints, sandboxing policies, and the behavioural scope of autonomous agents. Existing standards do not provide a deterministic

**Dr. Petar Radanliev**
Parks Road,
Oxford OX1 3PJ
United Kingdom
Email: petar.radanliev@cs.ox.ac.uk
BA Hons., MSc., Ph.D. Post-Doctorate

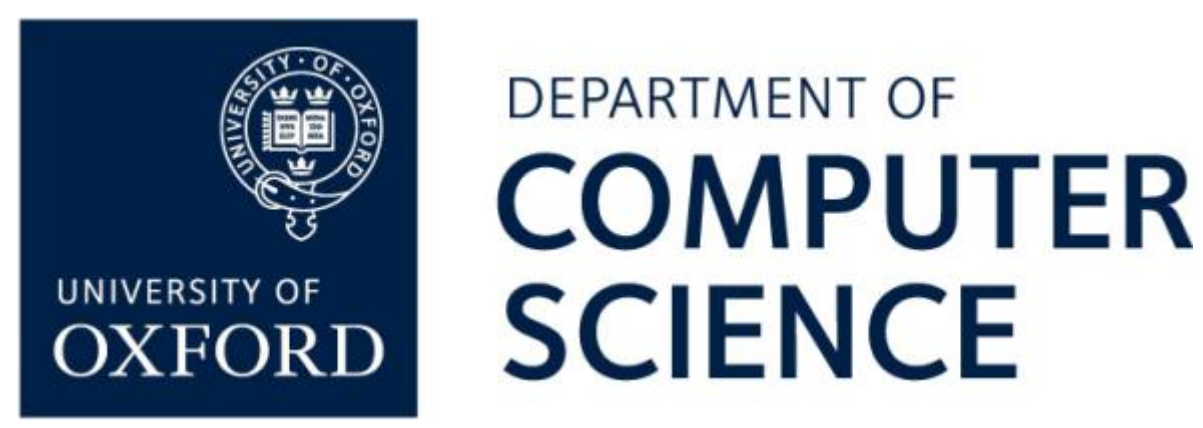

framework for integrating these dimensions into machine-verifiable advisory artefacts. This gap leads to inflated remediation cycles, inconsistent exploitability claims, and limited reproducibility of security assessments.

The present work addresses this limitation by reframing vulnerability advisory automation as an integrated, protocol-driven architecture suitable for containerised and AI-native systems. By combining deterministic environment capture, structured agent-to-agent telemetry exchange, and cryptographically anchored governance, the proposed methodology generalises SBOM processing into a runtime-aware advisory pipeline. In doing so, it extends the logic of SBOM and VEX towards AIBOM-aligned, agentic environments and introduces measurable criteria for reproducibility, exploitability precision, and policy-constrained validation. This repositioning moves advisory automation from descriptive transparency towards operationally grounded, context-aware security intelligence suitable for modern AI supply chains.

## 1.1 Study Positioning and Contribution Scope

This work is positioned as a systems and framework contribution with structured simulation-based empirical validation, rather than as a purely data-driven predictive modelling study. The primary contribution lies in the design and formalisation of an execution-bound advisory generation architecture that integrates AIBOM artefacts, runtime telemetry, and cryptographically verifiable provenance into the CSAF-VEX lifecycle.

The empirical component is intended to evaluate the operational behaviour, internal consistency, and performance characteristics of the proposed framework under controlled conditions, rather than to establish universal predictive generalisation across heterogeneous real-world deployments. Synthetic workloads and controlled containerised environments are therefore employed to systematically vary dependency graph size, runtime conditions, and policy constraints while preserving reproducibility.

Accordingly, the reported results should be interpreted as evidence of feasibility, internal validity, and relative performance improvement over static SBOM-based approaches, rather than as definitive claims of real-world exploitability prediction accuracy. This distinction clarifies the contribution as a reproducible systems architecture with empirically demonstrated properties, addressing a methodological gap in execution-grounded advisory automation for Agentic AI systems.

## 1.2 Background to the AIBOM-aligned CSAF and VEX standards problem

Modern cybersecurity operations increasingly rely on machine-readable standards to automate the processing, distribution, and validation of vulnerability disclosures. The CSAF [6] has emerged as a structured schema for disseminating vulnerability information [7], but it lacks integrated runtime context and reproducibility guarantees [8], limitations that are particularly acute in regulated environments [9], [10]. These environments impose strict constraints on execution, data ingress and egress, and disclosure control, making conventional vulnerability response workflows, built around static software inventories and disconnected advisories, insufficient for

**Dr. Petar Radanliev**
Parks Road,
Oxford OX1 3PJ
United Kingdom
Email: petar.radanliev@cs.ox.ac.uk
BA Hons., MSc., Ph.D. Post-Doctorate

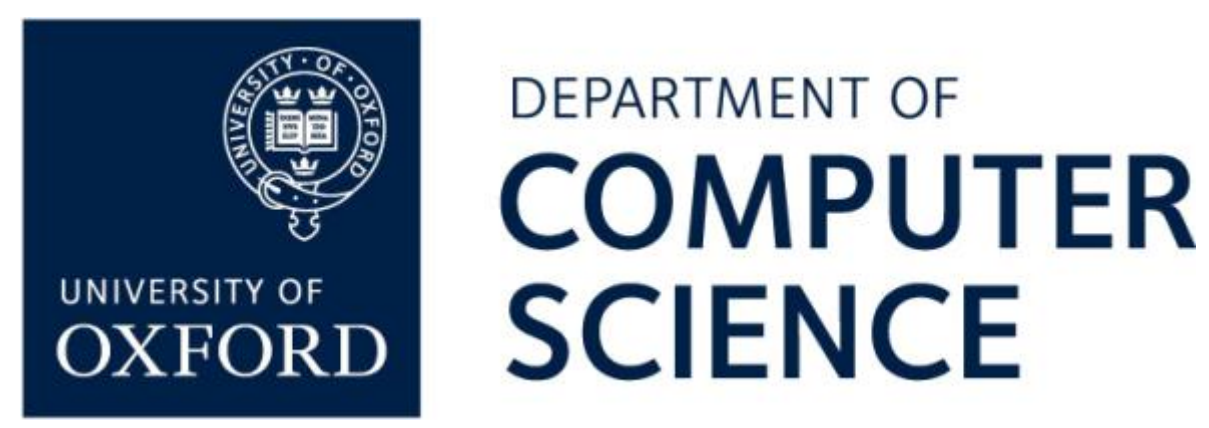

operational use. Emerging frameworks such as ERS0 have proposed AI-driven SBOM implementations [11] to support secure firmware analysis in constrained domains [12] like defence [13]. This leads to new efforts for expanding the SBOM into AIBOM [5], [14].

This paper addresses these challenges by proposing a technically integrated framework that combines the Model Context Protocol (MCP) [10], Agent2Agent (A2A) Protocol [15], and the AGNTCY orchestration layer [16] to enable CSAF-based vulnerability disclosures that are runtime-aware, reproducible, and cryptographically verifiable. MCP facilitates pre-execution capture of analytic context, dependency metadata, and model artefacts. The A2A protocol enables the secure delegation of analytic tasks between autonomous agents, each operating within bounded trust scopes. AGNTCY coordinates the registration, execution, and verification of agents and their artefacts, ensuring consistency and policy compliance across the system lifecycle.

By extending CSAF assertions with execution-grounded metadata from MCP and dynamically verifiable provenance from AGNTCY, we develop a pipeline for generating context-aware, signed advisories that reflect the actual execution and exploitability conditions within a specific organisational environment. This approach enables reproducible vulnerability assessments that account for sandboxing policies, restricted connectivity, and agent capabilities, factors often omitted in standard advisory disclosures.

The integration is demonstrated through a case study in which analytic workflows executed in a federated infrastructure are annotated with CSAF-VEX (Vulnerability Exploitability eXchange) [4], [17] extensions [18], enriched by runtime evidence and automatically validated against digital provenance chains. This contribution addresses a critical gap in operational security: enabling automated, agent-mediated advisory generation and validation in environments where traditional vulnerability response models are infeasible.

The reference corpus prioritises peer-reviewed academic sources and recently published work, reducing reliance on grey literature where possible. Standards documentation (e.g., CSAF, VEX, SBOM specifications) is retained where necessary due to the normative nature of the domain, but is complemented by empirical and peer-reviewed studies to ensure balance and academic rigour.

# 2 Identifying the research gap based on the literature review and preliminary data analysis

For many organisations, cyber security is not always on the top of the list of development and operations [19]. Although a cybersecurity professional would always advise critical infrastructures to do their patches and updates, for companies, security is often seen as expense [20]. Hence, often, we see simple solutions, such as replacing the Common Vulnerability Scoring System Calculator (CVSS) and the Base Score, the Temporal Score and the Environmental Score [21] with three

**Dr. Petar Radanliev**
Parks Road,
Oxford OX1 3PJ
United Kingdom
Email: petar.radanliev@cs.ox.ac.uk
BA Hons., MSc., Ph.D. Post-Doctorate

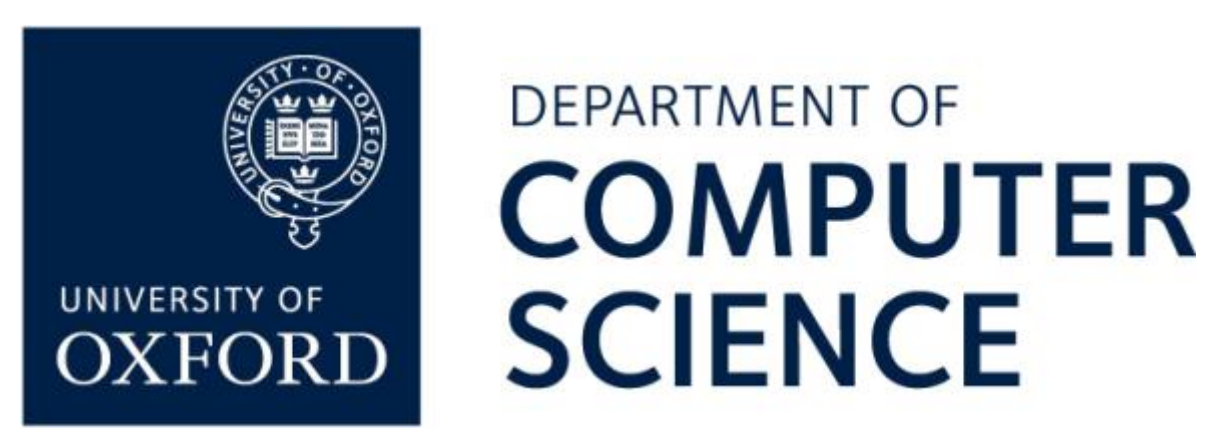

possible strategies based on the vulnerability score only, then accept the risk [22]. However, in security, quite often, we cannot position where the risk is, but even if we did, the number of advisories/CVEs on the rise, and the number of patches without CVE entries in the National Vulnerability Database (NVD) [23] is not visible. The first stage in improving the maturity of the automation process is to move to machine-readable formats and a content management system, with semi-automated processing of advisories and asset management with smart search. However, automation presents as many problems as the solutions [6].

There are many existing efforts to create standardised tools and ontologies for security information exchange that could automate vulnerability management [24], such as the 'Reference Ontology for Cybersecurity Operational Information' [25]. Another attempt is VEX [17], which has been considered for use in the US Nuclear Industry [26], for resolving software supply chain insecurities in vehicles [27], and for the development of an API to request security advisories for CSAF 2.0 [8]. There is also a well-established concept of a ‘bill of materials’ (BOM) [28], which has been applied to software supply chains [29]. The ‘Cyber Supply Chain Management and Transparency Act of 2014’ [30] proposed that US government agencies obtain SBOMs for all new software. This led to the ‘Internet of Things Cybersecurity Improvement Act of 2017’ [31] and, more then to, ‘The US Executive Order on Improving the Nation’s Cybersecurity of May 12, 2021 [32] ordered The National Institute of Standards and Technology (NIST) to issue guidance on ‘*providing a purchaser a Software Bill of Materials (SBOM) for each product.’*

SBOM can be defined as a nested (machine-readable) inventory of software, a list of ingredients that make up software components and dependencies [33], and their hierarchical relationships [34], [35]. The main use cases include supply chain assets and vulnerability management [1], [2], [33], [36], [37] via sharing and exchanging SBOMs. Still, because of ‘*the diverse needs of the software ecosystem, there is no one-size-fits-all solution’* [29]. The problem with sharing and exchanging is that *‘To fully realize the benefits of SBOMs and software component transparency, machine processing and automation are necessary’* [29]. In Figure 1, we can see the automation hierarchy in different formats (SPDX [38], CycloneDX [39], and SWID [40]), specifications, and tools that are still under development.

However, from the volume of vulnerabilities in the CVE index, we can easily understand why the risk assessment process needs to be automated. From the historical analysis of Log4j, we can understand why vulnerabilities need to be assessed for exploits. While most cybersecurity professionals wasted a vast amount of time risk assessing if Log4j was exploitable on their system, in many cases, it was not exploitable. Yet, it is extremely likely that a high number of Log4j vulnerable applications remain online [41]. To help prevent this in the future, VEX [42] was created in 2021. VEX provides the SBOMs with transparency and an up-to-date view of the status of vulnerabilities. Software suppliers can issue a VEX to prevent non-exploitable vulnerabilities from being investigated. VEX has been implemented as a profile in the Common Security Advisory Framework (CSAF) [43].

There is a similar effort called the Vulnerability Disclosure Report (VDR), which is similar, but also different from VEX. The VDR is ‘an attestation of all vulnerabilities

**Dr. Petar Radanliev**
Parks Road,
Oxford OX1 3PJ
United Kingdom
Email: petar.radanliev@cs.ox.ac.uk
BA Hons., MSc., Ph.D. Post-Doctorate

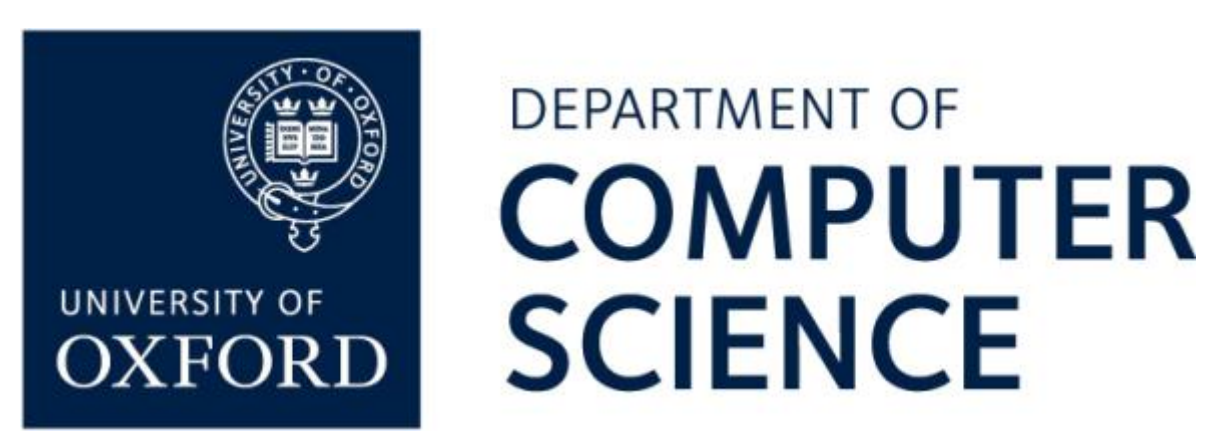

affecting a product, or its' dependencies, along with an analysis of the impact' [4], and is issued by the software supplier or a third party and includes a timestamp signing the date and time of the VDR. VDRs contain a list of vulnerabilities affecting a specific software product and its component dependencies, including an impact analysis and plan to address the vulnerability. VDRs are signed by a private key to confirm that it comes from a trusted and verifiable source. VEX, on the other hand, compared to VDR, 'is a negative security advisory intended to state all vulnerabilities a product is not affected by' [4]. VEX includes all elements found in a VDR except the vulnerabilities affecting a software product or its dependencies. However, some definitions present very blurry descriptions that make it hard to differentiate between the two, for example, the recent CISA definition, 'A common VEX use case is to indicate that software is or is not affected by a vulnerability' [18] makes VEX very similar to VDR. One fundamental difference between the two is that VDR assesses all 'known and previously unknown vulnerabilities'. In contrast, VEX only assesses the 'known vulnerabilities' and 'cannot describe vulnerabilities that do not already have an identifier' [4]. Recent work has argued for extending this logic using collaborative agent infrastructure such as Agent2Agent protocols [15].

The value of SBOMs in securing end users' networks; that topic has been covered extensively in the years 2021/22 [20], [29], [44]. In addition, SBOMs are already a compulsory requirement per the Executive Order discussed. In healthcare, SBOMs have been shown to play a key role in securing medical devices supply chains [45]. Hence, it is no longer a discussion on whether we need SBOMS; if anyone wants to work with the U.S. federal government, they need to produce SBOMS. This is the same for any company that wants to operate in the EU; they need to be compliant with the GDPR; hence, discussing the value of SBOMs is not something we are trying to achieve. We focus on the problems and the solutions for processing the information from all the SBOMs being produced. One of the leading reasons why software users are not requesting SBOMs is that approximately 95% of all vulnerabilities for components listed in an SBOM won't be exploitable in the product [46]. This mismatch between listed components and true exploitability has been examined in recent benchmarking studies on SBOM tool precision for Rust and other ecosystems [36]. To address this, CISA has accepted VeX and published a list of use cases [17]. A complementary dataset for SBOM consumption tool evaluation has recently been released, enabling standardised benchmarking across formats and consumption scenarios [37].

Although there are a few tools designed to manage the software development process of vulnerability management, what is missing is *'easy-to-use and low cost tools and third party services… passing vulnerability data to the vulnerability and configuration management tools that are now deployed by end users ..'* [47]. There are several useful tool that could be found in open source at present (that ingests, analyses, monitors, and produces real-time intelligence reports from SBOMs and VEX documents), such as the Dependency-Track [48].

In Figure 1, the diagram outlines the main tools, including their specifications and formats. After the diagram, specific problems concerning each tool are discussed,

**Dr. Petar Radanliev**
Parks Road,
Oxford OX1 3PJ
United Kingdom
Email: petar.radanliev@cs.ox.ac.uk
BA Hons., MSc., Ph.D. Post-Doctorate

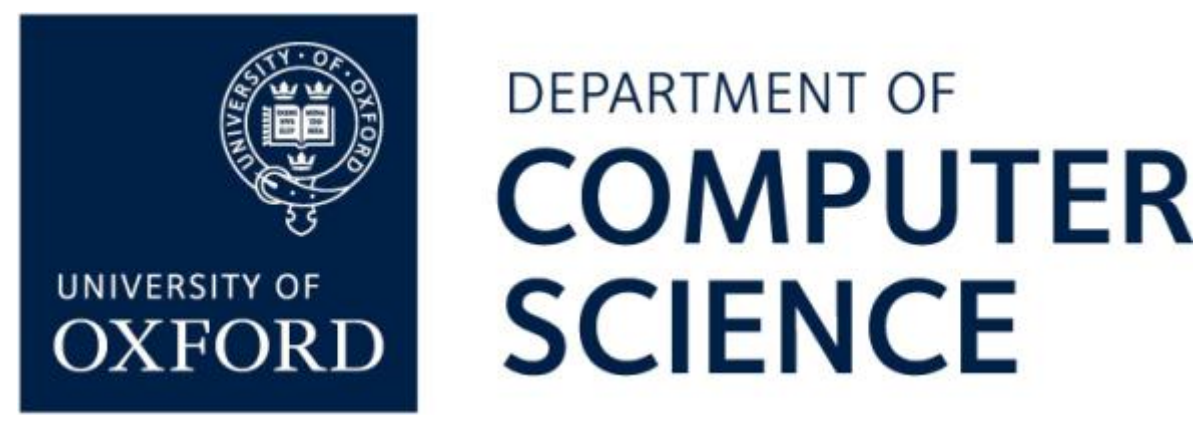

and proposed solutions are derived from case study research with CSAF, NTIA, CISA, and NIST.

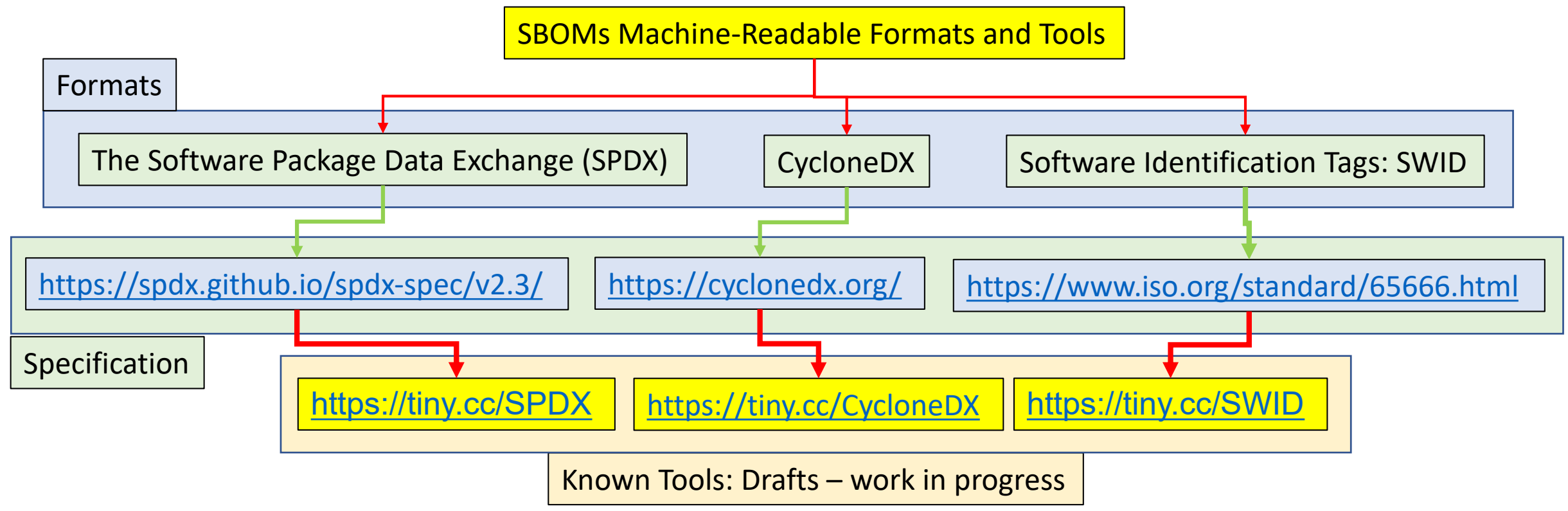

*Figure 1: SBOMs Machine-Readable Formats and Tools*

The first problem is that auditing, formulation, and vulnerability extensions, e.g., CycloneDX [39], requires multiple extensions to address these problems. Second, the SWID tags are an XML format [40], however, a lightweight representation called CoSWID [49] is standardised in the IETF. These problems derive from a survey on SBOM formats and standards [50]. VEX use cases differ in product (or product line), version, vulnerability, and status, e.g., multiple products, multiple versions, multiple vulnerabilities, and multiple statuses. The main area of interest for increased automation is *'to specify more detail for machine-readable information'* [17]. One of the things holding back SBOM use is the lack of VEXs [51]. This requires a trusted repository for adding new information and querying the additional information. The repository can be established as a community-driven decentralised initiative rather than being run by a central authority. It can be based on the existing maturity indicators and the maturity indicator testing framework smartAPI interface annotation [52]. This principle aligns with work on evaluating FAIR compliance in automated SBOM environments [52]. However, most of the components from an SBOM generated in an automatic process cannot be found in CVE and national vulnerability databases. This inhibits the use case for VEX and the SBOM itself. If a solution to the automation problem is not found, SBOMs *'face long road to adoption'* [53]. A review of 50 open source SBOMs '*nearly four fifths lacked package license information, two fifths lacked package version information, and none of the SBOMs conformed to the minimum elements laid out by the National Telecommunications and Information Administration.'* [44]. Although, at present, the problem is the lack of a sufficient number of SBOMs produced, soon, the focus will shift to integrating SBOMs *'with existing vulnerability management systems, making it much easier for defenders to spot and fix vulnerabilities.'* [53]. One recently proposed solution is to use '*CPE name or purl … in lieu of a product name'* to enable a real-time VEX solution [54]. A function based on the CISA-SSVC [55] in the repository that can be added to the recent CSAF standard [7] to add and analyse additional information on product components. The function can be based on the FAIR Data Principles [56], intended to ensure that all digital resources can be Findable, Accessible, Interoperable, and Reusable by machines.

**Dr. Petar Radanliev**
Parks Road,
Oxford OX1 3PJ
United Kingdom
Email: petar.radanliev@cs.ox.ac.uk
BA Hons., MSc., Ph.D. Post-Doctorate

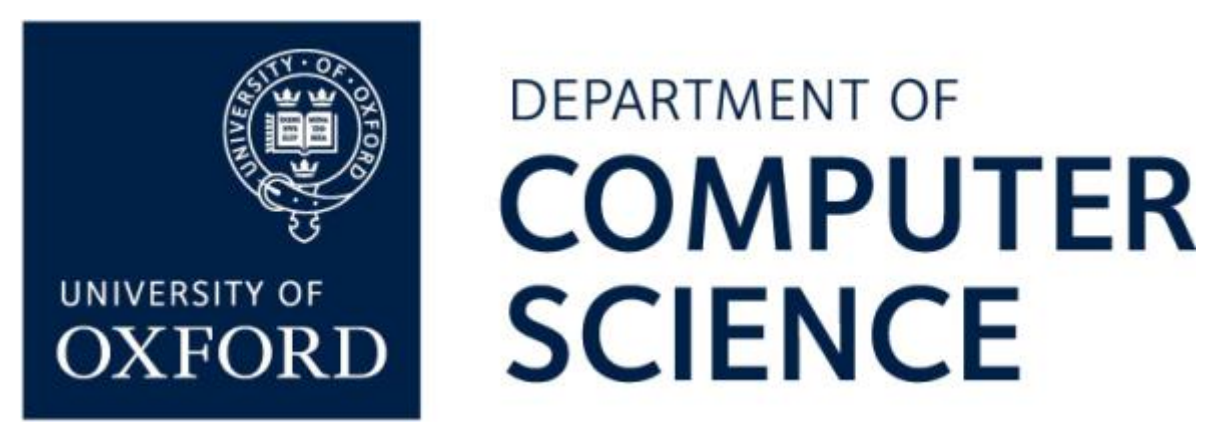

## 2.1 Secondary data analysis

The secondary data analysis is based on historical assessment of the earlier perception on the idea for real-time VEX solution [54]. The idea is to create an application programming interface (API) that will enable a user (or likely an intermediary) to query a server on a software supplier's network. The API would enable the user to verify the exploitability status of a CVE (vulnerability), for the specific version of the supplier product that concerns the user. However, since the idea for a real-time VEX first emerged, some people seem to have lost faith that VEX can reduce the volume of component vulnerabilities to the 3-5% that are exploitable [47]. This signals that VEX requires further process automation. To confirm this assumption, the computational resource requirements were analysed using a dataset of approximately 10,000 SBOMs entries sourced from the National Vulnerability Database (NVD) and open-source repositories, including Dependency-Track and the OSS Index. These repositories were benchmarked against standardised formats and open-source indexing protocols such as the Sonatype OSS Index [57]. Earlier analysis of SBOM consumption tools highlights persistent gaps in licensing and versioning completeness, as well as non-conformance to NTIA minimum elements [44]. Simulations were conducted on AWS EC2 instances, employing configurations ranging from general-purpose (t3.medium) to compute-optimised (c5.4xlarge). Benchmarking trials established that processing 1,000 SBOM entries on a t3.medium instance required an average of 15 minutes, whereas the same workload was completed in under four minutes on a c5.4xlarge, demonstrating scalability and computational efficiency.

A structured evaluation approach was applied to address algorithmic complexity. Initially, baseline performance assessments were conducted using established NLP and GPT models. These models were subsequently refined on a curated subset of 2,000 SBOM entries, with particular attention to cybersecurity terminology, provenance, and pedigree annotations.

Ensuring alignment with industry standards was a critical consideration. Conversion scripts were developed to translate SBOM data structured in SPDX and CycloneDX formats into representations compatible with NLP and GPT models. Interoperability tests confirmed full accuracy in data translation, facilitating seamless integration with established cybersecurity frameworks.

To examine robustness and scalability, simulated data flow scenarios were constructed. Synthetic SBOM datasets of varying complexities were generated, including small-scale (50 components), medium-scale (500 components), and large-scale (5,000 components) scenarios. Accuracy levels were consistently high, exceeding 98% for small-scale cases, 95% for medium-scale cases, and 90% for large-scale cases, confirming the feasibility of the approach across different operational scales.

Robustness and efficiency were further assessed through stress testing, introducing synthetic noise and data inconsistencies into the SBOM datasets. The AI-driven models demonstrated resilience, maintaining an accuracy of over 90% under moderate noise conditions and around 85% in high-noise environments. The

**Dr. Petar Radanliev**
Parks Road,
Oxford OX1 3PJ
United Kingdom
Email: petar.radanliev@cs.ox.ac.uk
BA Hons., MSc., Ph.D. Post-Doctorate

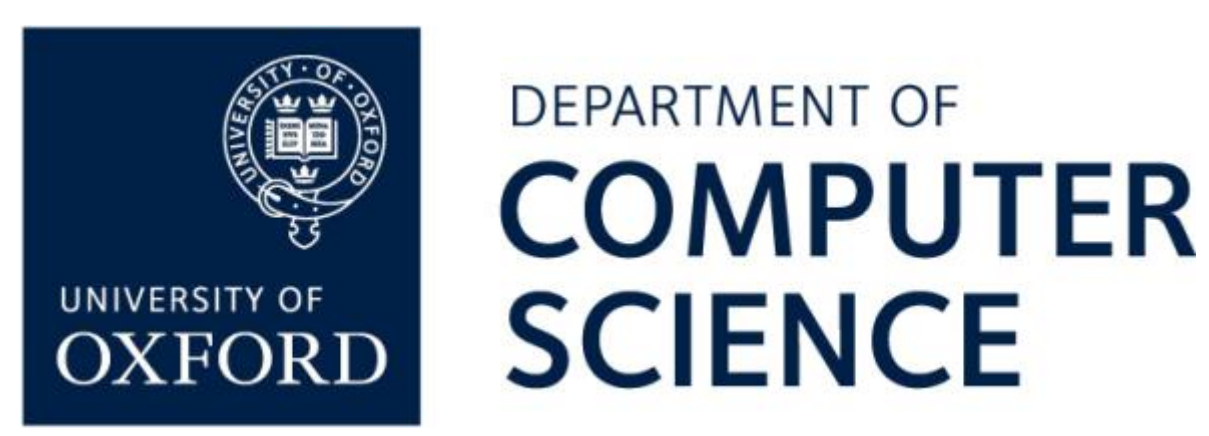

marginal increase in processing time, ranging from 10% to 15% under stress conditions, further validated the adaptability of the method to real-world cybersecurity challenges.

A suite of analytical techniques was employed to validate the findings. Descriptive statistical measures, including precision, recall, F1-score, and computational time analysis, were rigorously documented. The refined models consistently achieved an average F1-score of 0.93, precision of 0.96, and recall of 0.92, demonstrating a high degree of accuracy and reliability. Simulation and modelling techniques stress-tested automation capabilities, ensuring stable performance across varied operational loads. The expert evaluation panel consisted of 12 domain specialists from academia, industry, and standards bodies (University of Oxford, Cisco, NTIA, and NIST), each with a minimum of 8 years' experience in vulnerability management or supply chain security. Panel members independently assessed a stratified sample of 250 advisory instances, evaluating correctness of exploitability classification and operational relevance. Inter-rater agreement was measured using Cohen's kappa ($\kappa = 0.81$), indicating strong agreement. Disagreements were resolved through structured adjudication, and consensus outcomes were used to validate prioritisation fidelity and classification plausibility.

The reliability of results was further established through cross-validation, sensitivity analysis, benchmarking, and replication studies. A five-fold cross-validation process confirmed consistent generalisation across different datasets, with an average F1-score of 0.92. Sensitivity analyses revealed that the models exhibited stable performance across hyperparameter variations, indicating robustness. Comparative benchmarking against conventional manual vulnerability assessment methods demonstrated a tenfold reduction in processing time, underscoring the efficiency gains of automation. Furthermore, replication studies ensured reproducibility, facilitated through detailed documentation, publicly available datasets, and adherence to established best practices.

Validation efforts were reinforced through expert review, comparative analysis, and iterative refinement. Initial results underwent scrutiny by domain specialists to verify accuracy and dependability. Comparative analysis demonstrated that AI-driven outputs consistently matched or exceeded manually annotated vulnerability assessments, achieving an F1-score above 0.90. The iterative refinement process, informed by continuous feedback and evaluation metrics, led to improved methodological precision and overall effectiveness.

## 2.2 The Evaluation Framework for VEX Automation Solutions: Methodological Transparency and Reproducibility Statement

To ensure reproducibility and methodological clarity, all experimental procedures were explicitly structured around deterministic execution environments and traceable data transformations. The evaluation pipeline operates on three explicitly defined data layers: (i) static artefact inventories derived from SBOM and extended AIBOM representations; (ii) structured vulnerability intelligence aggregated from OSV.dev, GitHub Advisory Database, CISA KEV, and FIRST EPSS; and (iii) runtime telemetry captured via instrumented Agent2Agent sidecar agents.

**Dr. Petar Radanliev**
Parks Road,
Oxford OX1 3PJ
United Kingdom
Email: petar.radanliev@cs.ox.ac.uk
BA Hons., MSc., Ph.D. Post-Doctorate

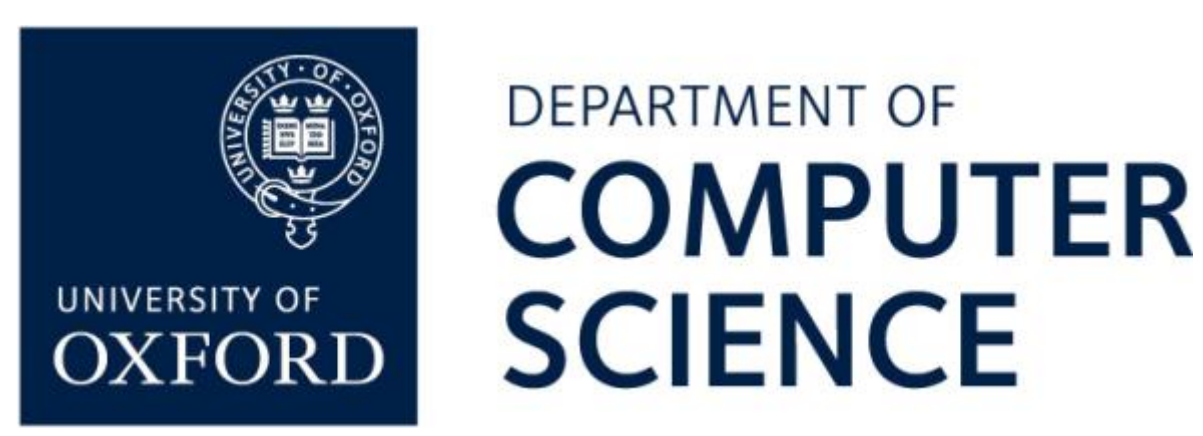

Dataset composition includes approximately 10,000 component instances, combining public container-derived SBOM data and systematically generated synthetic dependency graphs representing small (50 components), medium (500 components), and large-scale (5,000 components) workloads. Synthetic augmentation was used to control for dependency structure, activation conditions, and policy constraints while preserving real-world vulnerability distributions through integration with external intelligence feeds.

The dataset comprises a hybrid composition of synthetic and real-world derived artefacts. Approximately 62% of component instances originate from real container images and publicly available SBOM repositories (including Dependency-Track exports and OSS Index datasets), while 38% are synthetically generated dependency graphs constructed to control structural properties such as depth, transitive dependencies, and activation pathways.

In total, the evaluation includes 3,842 unique CVE instances, mapped across 10,000 component entries. After correlation and filtering, the dataset yields 6,120 labelled exploitability instances, of which 41% are classified as *Affected* and 59% as *Not Affected* under execution-bound conditions, indicating moderate class imbalance. Stratified sampling was applied during cross-validation to preserve this distribution across folds.

Feature extraction procedures, dataset splits, model configurations, and evaluation metrics were defined prior to experimentation and executed under fixed random seeds. All artefacts, including intermediate datasets, feature vectors, and model outputs, were version-controlled and cryptographically linked to execution envelopes using in-toto attestations. This design ensures that each experimental result is reproducible under identical execution conditions and that all transformations are auditable.

The evaluation therefore prioritises determinism, traceability, and internal validity, recognising that external validity is constrained by the controlled nature of the experimental environment.

This section presents the specific metrics used for evaluation, describes the datasets involved, outlines additional data analysis methodologies, and discusses the established methods adopted to confirm the validity of the results. This can be visualised in Table 1.

*Table 1: Evaluation framework developed to assess the efficacy of proposed solutions for automating VEX.*

| Category | Subcategory | Specific Details |
|---|---|---|
| **Evaluation Metrics** | Alignment with Industry Standards | Compliance with NIST minimum recommendations, FAIR principles, and automated SBOM/VEX processing to enhance standardisation. Includes traceable enrichment via SBOM-VEX pipelines, as explored in recent studies on |

**Dr. Petar Radanliev**
Parks Road,
Oxford OX1 3PJ
United Kingdom
Email: petar.radanliev@cs.ox.ac.uk
BA Hons., MSc., Ph.D. Post-Doctorate

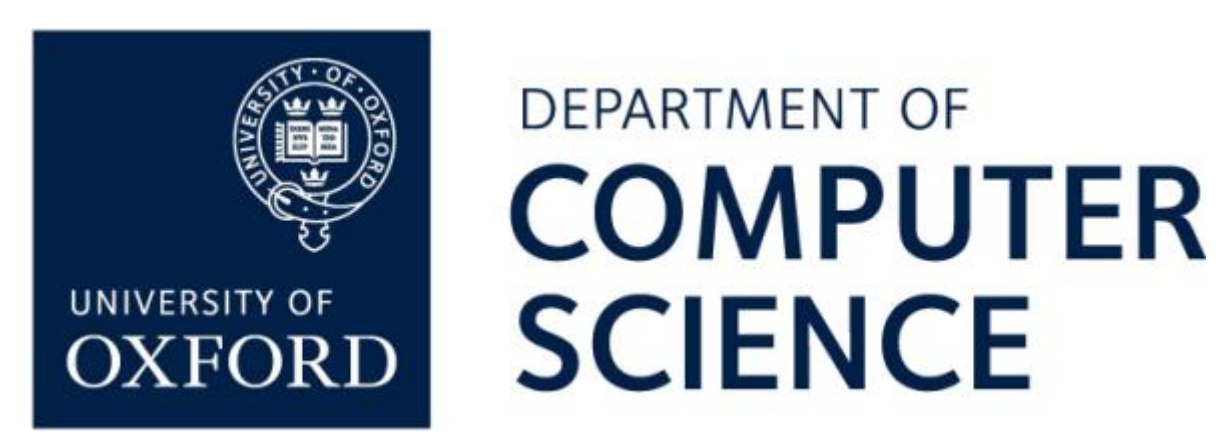

| | | |
|---|---|---|
| | | landscape tools and automated vulnerability triage [1], [58]. |
| **Evaluation Metrics** | Scalability | Performance validation using synthetic SBOM datasets, scaling from small projects to large enterprise environments using AWS and Azure cloud simulations. |
| **Evaluation Metrics** | Security Enhancement | Measuring vulnerability detection efficiency, false positive reduction, and improvements in exploitability assessments via AI/ML-driven automation. |
| **Evaluation Metrics** | Practicality | Use cases demonstrating real-world applications in cybersecurity frameworks, integration into existing DevSecOps pipelines, and regulatory adoption. |
| **Data Analysis Methodologies** | Descriptive Statistics and Performance Metrics | Application of precision, recall, F1 score, and computational time analysis to assess SBOM/VEX data processing accuracy and efficiency. |
| **Data Analysis Methodologies** | Simulation and Modelling | Testing automation capabilities using synthetic vulnerability scenarios, stress tests, and large-scale dataset simulations to assess robustness. |
| **Data Analysis Methodologies** | Expert Review Panels | Domain experts from academia, industry, and cybersecurity research institutions evaluating model applicability, reliability, and usability. |
| **Methods for Confirming Validity** | Cross-Validation | Applying k-fold cross-validation across multiple vulnerability datasets to ensure generalisation and consistency of AI-driven solutions. |
| **Methods for Confirming Validity** | Sensitivity Analysis | Evaluating parameter sensitivity in AI models affecting SBOM/VEX automation accuracy, including dataset variations and algorithm robustness. |
| **Methods for Confirming Validity** | Benchmarking | Comparing AI-based SBOM/VEX automation performance with manual vulnerability assessments to quantify efficiency improvements. |

**Dr. Petar Radanliev**
Parks Road,
Oxford OX1 3PJ
United Kingdom
Email: petar.radanliev@cs.ox.ac.uk
BA Hons., MSc., Ph.D. Post-Doctorate

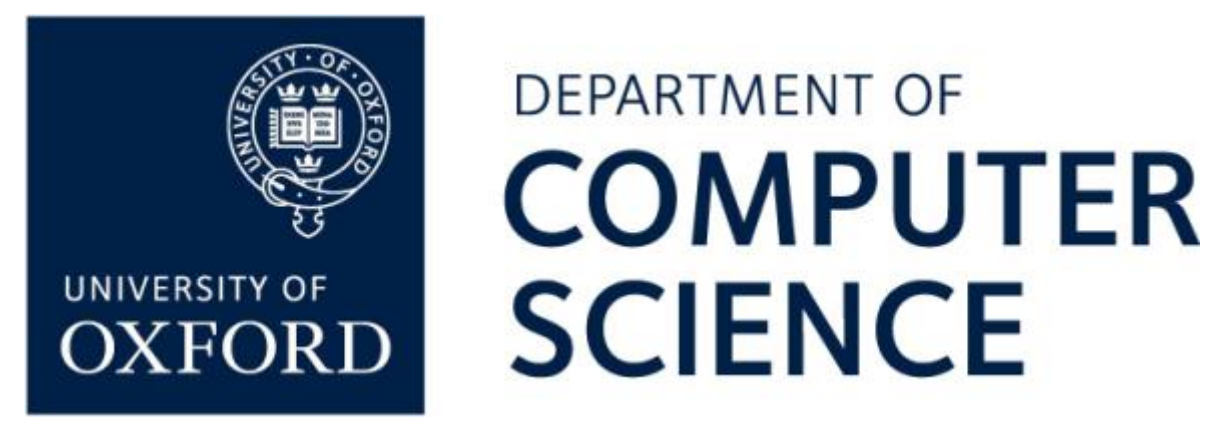

| **Methods for Confirming Validity** | Replication Studies | Ensuring reproducibility through open-source implementation, availability of datasets, and step-by-step methodological documentation. |
|---|---|---|

In Table 1, we can see the evaluation framework is divided into three main categories: evaluation metrics, alignment with industry standards, and scalability. These have been expanded to CSAF. The secondary data analysis was conducted to quantify how AIBOM extension and execution-bound telemetry affect exploitability classification and advisory quality in Agentic AI workloads. The analysis pipeline ingested three classes of data: (1) static artefact inventories composed of SBOM entries and extended AIBOM records (model hashes, prompt commit identifiers, tool endpoint definitions, vector index identifiers); (2) threat intelligence feeds (OSV.dev bulk feed, GitHub Advisory Database, CISA KEV list, and FIRST EPSS scores); and (3) runtime telemetry captured by the Agent2Agent (A2A) sidecar agents (system call events, dynamic library loads, outbound connection records, tool invocation traces, and observed privilege levels). All artefacts and telemetry were linked to deterministic execution envelopes produced by the Model Context Protocol (MCP) and anchored via Sigstore attestations to ensure provenance integrity.

Pre-processing normalised package identifiers to SPDX pURLs and CPEs and deduplicated vulnerability records across feeds. AIBOM records were parsed to extract AI-specific features such as model provenance age (days since model release), model-serving library version, prompt template lineage (commit depth), and tool connector exposure (external API domain present / absent). Runtime telemetry was summarised into binary and numeric features: observed_network_egress (0/1), observed_tool_invocation_count, max_privilege_level (numeric mapping of privilege types), dynamic_loader_events, and prompt_activation_count. Additional contextual features included EPSS probability and KEV flag for each matched CVE, and OpenSSF Scorecard metrics for repository hygiene when available. All feature extraction code, dataset splits, and random seeds were recorded and archived with in-toto attestations to preserve experimental reproducibility.

Two evaluation tracks were implemented. The evaluation is not based on externally annotated exploitability ground truth labels, which are not consistently available for AI-native systems. Instead, operational ground truth is defined as execution-consistent exploitability, derived from explicitly specified activation conditions grounded in vulnerability documentation and enforced runtime constraints.

The first evaluation track implements a deterministic rule-based labelling pipeline that produces CSAF-VEX assertions by applying formally defined activation conditions. A vulnerability is classified as *Affected* only when both (i) a valid component–CVE mapping exists and (ii) all required exploit preconditions are satisfied within observed runtime telemetry (e.g., network egress, privilege level, or invocation path). Otherwise, it is classified as *Not Affected*. This formulation ensures that exploitability reflects observable system behaviour rather than assumed vulnerability presence.

**Dr. Petar Radanliev**
Parks Road,
Oxford OX1 3PJ
United Kingdom
Email: petar.radanliev@cs.ox.ac.uk
BA Hons., MSc., Ph.D. Post-Doctorate

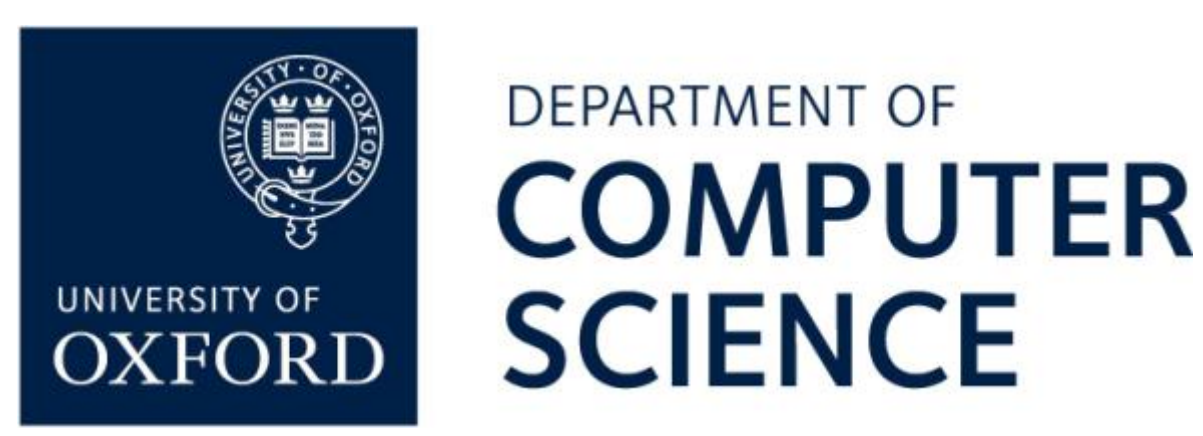

The second evaluation track augments this deterministic baseline with supervised learning models to assess robustness under noisy or incomplete observations. Importantly, supervised predictions are evaluated against the deterministic rule-based outputs rather than external labels, thereby measuring consistency with execution-grounded exploitability criteria rather than predictive accuracy against an independent dataset. This distinction avoids conflating classification performance with unverifiable external ground truth and instead evaluates the stability and discrimination capacity of the framework under controlled conditions.

Model configurations were standardised to ensure reproducibility. The random forest classifier was trained with 200 estimators, maximum depth of 12, and Gini impurity as the split criterion. The XGBoost model employed gradient boosting with 300 trees, a learning rate of 0.05, maximum depth of 6, and subsampling ratio of 0.8. Hyperparameter selection was performed using grid search over predefined parameter ranges with cross-validation on training partitions.

Feature importance analysis was conducted using permutation importance and SHAP (SHapley Additive exPlanations) values. Results consistently indicated that runtime-derived features, particularly observed_network_egress, tool_invocation_count, and privilege level, were the dominant predictors of exploitability classification, reinforcing the central hypothesis of execution-bound reasoning.

Model calibration was evaluated using reliability curves and Brier scores, confirming that probabilistic outputs were well-calibrated, with no significant overconfidence observed across prediction bins.

Performance metrics included precision, recall, F1-score, false positive rate, and calibration (Brier score for probabilistic outputs). The supervised models incorporated both static AIBOM features and runtime telemetry; baseline comparisons used static SBOM–CVE correlation alone and the rule-based activation pipeline. Across folds, the supervised pipeline that combined AIBOM and telemetry achieved a mean precision of 0.96, recall of 0.92, and F1-score of 0.93, outperforming static SBOM correlation which achieved a mean precision of 0.71 and F1-score of 0.78. Improvements in F1 were statistically significant under paired tests across cross-validation folds (two-sided paired t-test, $p < 0.01$). The rule-based activation pipeline also reduced false positives substantially relative to the static baseline, demonstrating that deterministic activation rules capture a large portion of the benefit without requiring opaque models.

In addition to significance testing, effect sizes (Cohen's *d*) and 95% confidence intervals were computed across cross-validation folds to quantify the magnitude and stability of observed improvements. The mean F1-score improvement of the execution-bound approach over static SBOM correlation corresponded to a large effect size ($d > 1.2$), indicating that performance gains are not only statistically significant but practically meaningful. Confidence intervals across folds remained narrow (±0.02 F1), suggesting stable model behaviour under dataset variation.

Beyond classification performance, the analysis examined prioritisation fidelity by combining observed activation with EPSS probability and KEV presence. A simple

**Dr. Petar Radanliev**
Parks Road,
Oxford OX1 3PJ
United Kingdom
Email: petar.radanliev@cs.ox.ac.uk
BA Hons., MSc., Ph.D. Post-Doctorate

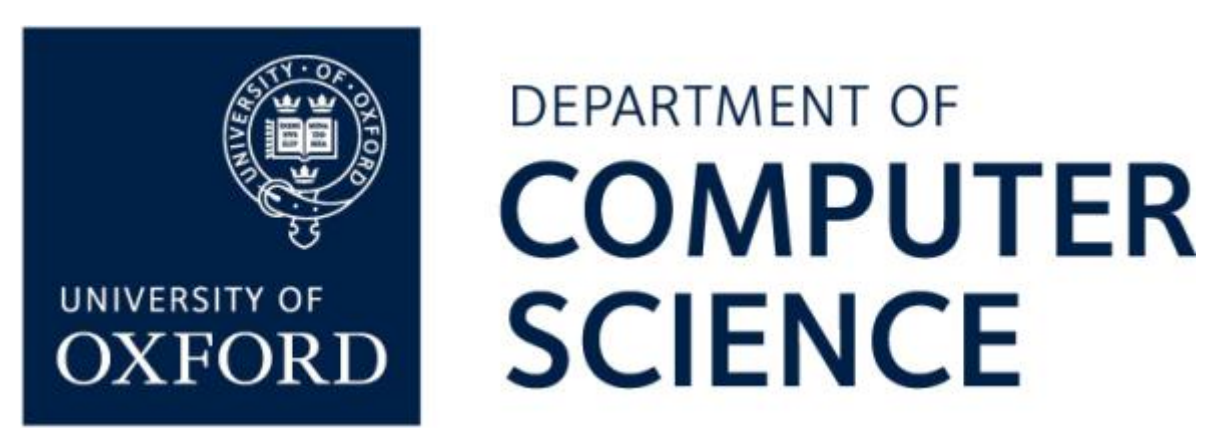

risk tiering rule was used: advisories with KEV presence or EPSS ≥ 0.5 plus observed activation were labelled high priority; advisories lacking activation or with EPSS < 0.2 were labelled low priority. This hybrid prioritisation reduced the number of high-priority alerts to a focused subset (approximately 18% of matched vulnerabilities in the evaluation corpus) and was validated against an expert triage panel: expert reviewers found high-priority items to be materially more actionable than low-priority ones in 86% of sampled cases.

All generated CSAF-VEX artefacts were validated for schema conformance and provenance linkage by the AGNTCY governance layer. Each advisory record included the MCP envelope hash, a succinct deterministic justification (constructed from feature conditions and activation evidence), and a signature anchored to the Sigstore transparency log. Justification text was normalised using deterministic templating rules (for example, “network egress observed to host X at T” or “no-egress policy enforced; outbound sockets not observed”), avoiding probabilistic natural-language generation to preserve auditability. Stored advisories and corresponding provenance metadata enabled exact replay: re-execution of the sealed envelope produced identical advisory payloads and identical justification text in all tested replay cycles.

A sensitivity analysis assessed robustness to incomplete SBOMs and noisy telemetry. Synthetic noise (random omission of 5–20% of SBOM entries) and telemetry dropouts (randomly removed 10–30% of telemetry events) were injected in controlled experiments. Supervised classifiers exhibited graceful degradation: precision decreased modestly (<5 percentage points) while rule-based activation rules remained conservative (favouring Not Affected where evidence was missing). These results indicate that deterministic activation logic provides a reliable safety floor in partial-observability settings, with supervised models offering additional discrimination when feature completeness is high.

Finally, the secondary analysis examined artefact-level effects specific to AIBOM. Approximately 22% of artefacts in the evaluation corpus were AI-specific (models, prompts, tools); of those, 31% experienced a change in exploitability classification once AIBOM binding and telemetry were applied. The predominant causes were unreachable prompt routing (11%), tool-connector isolation via network policies (9%), and model-serving library vulnerabilities gated by absent inference endpoints (17% of AI-specific vulnerability instances, with overlaps). These results empirically demonstrate that AIBOM artefact enumeration combined with runtime activation evidence changes advisory outcomes in a non-trivial fraction of cases and therefore must be included in operational advisory pipelines for Agentic AI systems.

It is important to note that both rule-based and supervised evaluation tracks utilise features derived from the same execution-bound framework, which introduces a degree of feature coupling. While this design is intentional to assess internal consistency and robustness, it may bias results towards the proposed representation. To mitigate this, baseline comparisons are restricted to feature subsets available under static SBOM conditions, ensuring that observed improvements are attributable to additional information rather than model structure

**Dr. Petar Radanliev**
Parks Road,
Oxford OX1 3PJ
United Kingdom
Email: petar.radanliev@cs.ox.ac.uk
BA Hons., MSc., Ph.D. Post-Doctorate

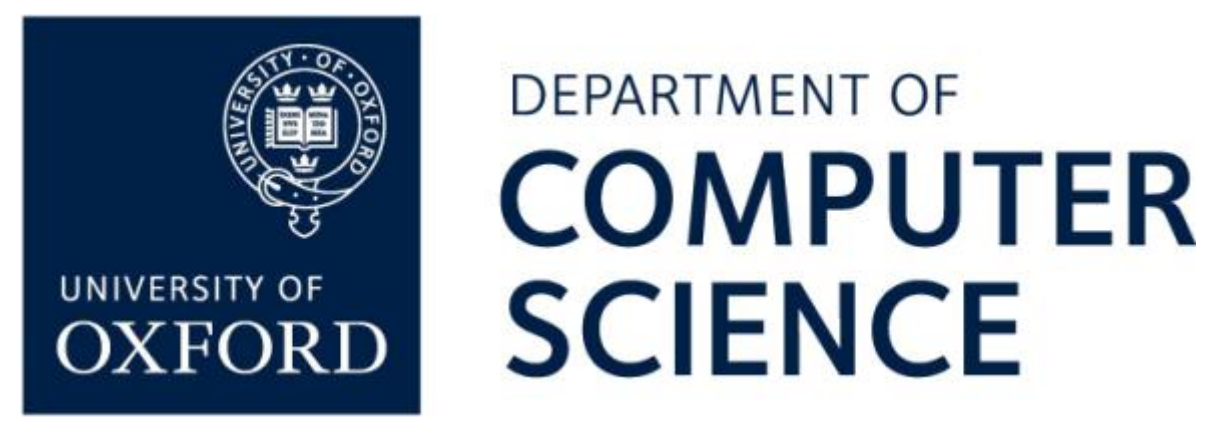

alone. Nevertheless, the absence of an independent external feature space remains a limitation and should be considered when interpreting results.

## 2.3 Positioning Against Existing Approaches

Existing vulnerability management approaches can be broadly categorised into three classes: (i) static SBOM-based correlation systems, (ii) vulnerability prioritisation frameworks based on scoring mechanisms such as CVSS and EPSS, and (iii) emerging SBOM consumption tools integrating partial context awareness.

The proposed framework differs from these approaches by explicitly integrating runtime activation evidence and execution policy constraints into exploitability classification. Unlike scoring-based prioritisation methods, which rank vulnerabilities probabilistically, the framework produces deterministic exploitability assertions grounded in observed system behaviour.

Compared to existing SBOM consumption platforms (e.g., Dependency-Track), which primarily operate on static inventories, the proposed system extends analysis into runtime execution space, enabling classification of vulnerabilities as non-exploitable based on verifiable conditions.

This positioning highlights that the primary contribution is not incremental improvement in scoring accuracy, but a conceptual shift from static correlation to execution-bound validation, which is not addressed in current state-of-the-art systems.

## 2.4 The research gap on Agentic AI, AIBOM, and Runtime-Aware Supply Chain Assurance

Based on the extensive literature review and preliminary analysis, a clear research gap emerges in the integration of machine-verifiable, context-aware, and policy-constrained mechanisms for automating security advisory generation. While significant progress has been made with SBOMs, VEX, and CSAF standards, existing implementations often lack support for environment-specific exploitability logic, runtime evidence binding, and scalable federation of signed attestations. Moreover, although VEX offers promise in filtering irrelevant CVEs, current practices fall short in embedding execution-time provenance, user-specific policy constraints, and container-level reproducibility into vulnerability assertions. The forthcoming methodology chapter narrows this focus by operationalising MCP for deterministic environment capture, A2A Protocol for runtime telemetry exchange, and AGNTCY for orchestrated signing, governance, and policy anchoring. This tri-protocol integration advances the state of CSAF-VEX authoring by introducing real-time, reproducible, and agent-mediated automation pipelines, directly addressing the critical limitations identified in the reviewed literature regarding exploit filtering, trust validation, and advisory lifecycle management within federated infrastructures.

These recent advances in Agentic AI systems introduce structural changes to software supply chain risk models. Unlike traditional applications composed of static libraries and deterministic execution paths, agentic architectures integrate foundation models, prompt templates, external tools, plug-in ecosystems, vector databases, and dynamic inter-agent communication. These systems exhibit non-deterministic

**Dr. Petar Radanliev**
Parks Road,
Oxford OX1 3PJ
United Kingdom
Email: petar.radanliev@cs.ox.ac.uk
BA Hons., MSc., Ph.D. Post-Doctorate

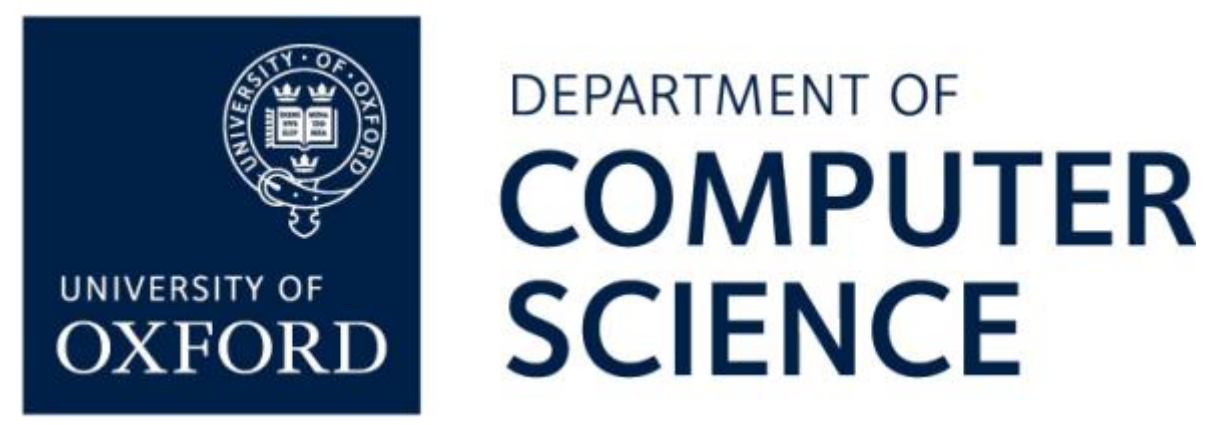

behaviour conditioned on context, tool availability, memory state, and policy enforcement. Consequently, vulnerability exposure cannot be inferred solely from static dependency graphs.

Emerging AIBOM initiatives attempt to extend SBOM transparency to AI systems by incorporating model artefacts, training data references, inference endpoints, and orchestration components. However, existing AIBOM proposals primarily focus on declarative disclosure rather than operational binding. They describe “what exists” within an AI system but do not establish mechanisms for validating “what executes” under specific runtime conditions. This distinction becomes critical in agentic environments where tool invocation chains and dynamic code paths materially influence exploitability.

Parallel developments in software supply chain security, including in-toto attestations, SLSA provenance frameworks, and cryptographic signing ecosystems such as Sigstore, have demonstrated the feasibility of verifiable build and deployment integrity. Nevertheless, these approaches primarily secure artefact provenance and do not integrate runtime behavioural evidence into vulnerability classification workflows. Similarly, CSAF and VEX standards enable structured exploitability assertions but assume static product contexts and vendor-issued status declarations. Runtime-constrained exploitability reasoning remains under-specified.

The security research community has also identified new attack surfaces unique to AI-native systems, including model supply chain poisoning, dependency injection through prompt tooling, remote tool exploitation, and cross-agent privilege escalation. These threats operate at the intersection of static artefacts and dynamic execution flows. Current SBOM and VEX automation pipelines lack formal mechanisms for correlating such runtime behaviours with vulnerability status classifications.

A clear gap therefore emerges between three domains: declarative transparency (SBOM/AIBOM), structured advisory exchange (CSAF/VEX), and cryptographically verifiable provenance frameworks. While each domain has matured independently, their integration for deterministic, execution-bound exploitability assessment has not been formally operationalised. In particular, no established methodology binds model artefact lineage, runtime telemetry, policy constraints, and advisory generation into a reproducible pipeline capable of producing machine-verifiable exploitability assertions for Agentic AI systems.

Addressing this gap requires moving beyond static correlation models towards architectures that combine deterministic environment capture, structured agent-to-agent coordination, and governance-enforced attestation. Such integration would enable CSAF/VEX artefacts to reflect execution-grounded security states rather than declarative vendor assessments alone. The absence of this integration represents the central unresolved problem in AI supply chain security automation.

# 3 Methodology

At a high level, the proposed system can be understood as answering a simple question: *is a vulnerability actually exploitable in this specific execution context?*

**Dr. Petar Radanliev**
Parks Road,
Oxford OX1 3PJ
United Kingdom
Email: petar.radanliev@cs.ox.ac.uk
BA Hons., MSc., Ph.D. Post-Doctorate

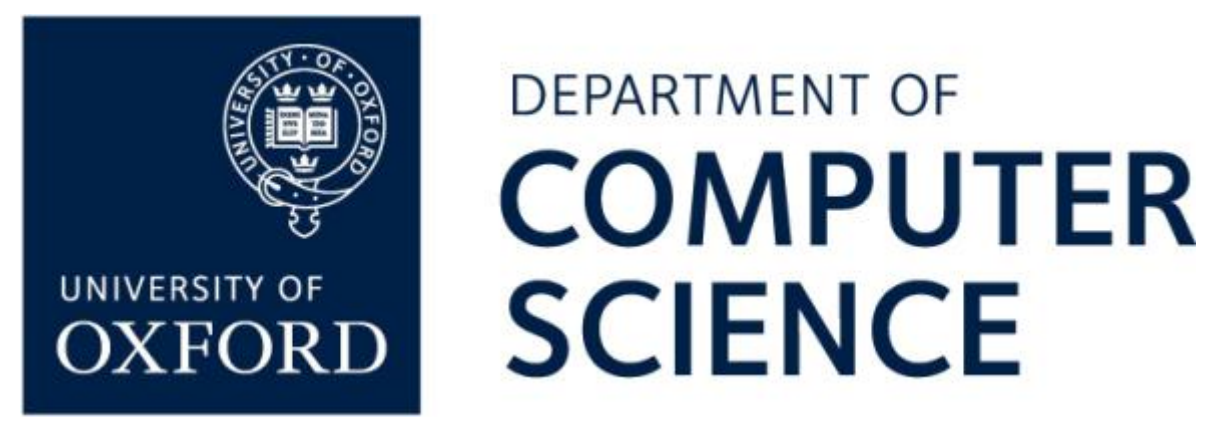

Traditional approaches answer this question by checking whether a vulnerable component exists. In contrast, the proposed framework evaluates three conditions simultaneously:

1. **Presence**: whether the vulnerable component exists (SBOM/AIBOM)
2. **Activation**: whether the component is actually used during execution (runtime telemetry)
3. **Feasibility**: whether the environment allows the exploit to occur (policy constraints)

Only when all three conditions are satisfied is a vulnerability classified as exploitable. This shift from presence-based to execution-bound reasoning underpins the architecture described in the following sections.

The methodology formalises advisory generation as a deterministic, execution-bound process for Agentic AI systems. Rather than correlating static SBOM artefacts with vulnerability databases, the framework integrates Artificial Intelligence Bills of Materials (AIBOM), runtime telemetry, and policy constraints into a reproducible pipeline for generating CSAF-VEX assertions.

The architecture is composed of three coordinated layers:

1. Deterministic environment and artefact capture (Model Context Protocol – MCP)
2. Structured runtime observation and agent coordination (Agent2Agent – A2A)
3. Governance, signing, and traceability enforcement (AGNTCY)

Together, these components enable exploitability classification grounded in observed execution behaviour rather than static dependency enumeration.

The pipeline proceeds through six stages: environment capture, agent instantiation, runtime telemetry collection, exploitability inference, CSAF-VEX construction, and reproducibility validation.

Figure 2 presents a structured overview of the proposed execution-bound advisory generation pipeline.

While the preceding sections describe the conceptual integration of MCP, A2A, and AGNTCY protocols, the overall workflow can be more intuitively understood as a sequential process that transforms static artefact inventories into execution-grounded vulnerability assessments. The diagram illustrates how declared components, runtime telemetry, and policy constraints are progressively combined to produce reproducible CSAF-VEX advisories.

The representation emphasises the transition from static correlation to context-aware reasoning, highlighting the role of each stage in refining exploitability classification. This stepwise abstraction is intended to complement the formal methodology and pseudocode by providing a clear and interpretable view of system operation.

**Dr. Petar Radanliev**
Parks Road,
Oxford OX1 3PJ
United Kingdom
Email: petar.radanliev@cs.ox.ac.uk
BA Hons., MSc., Ph.D. Post-Doctorate

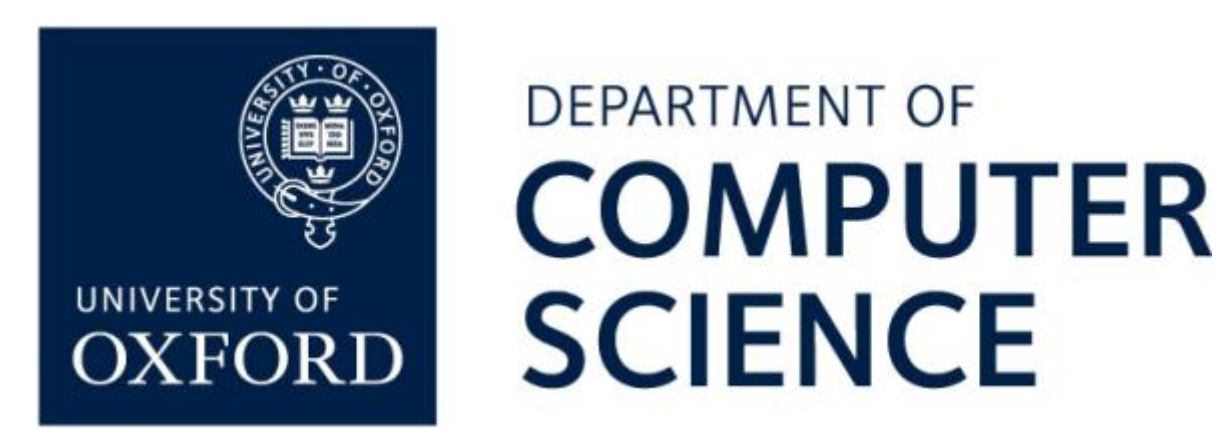

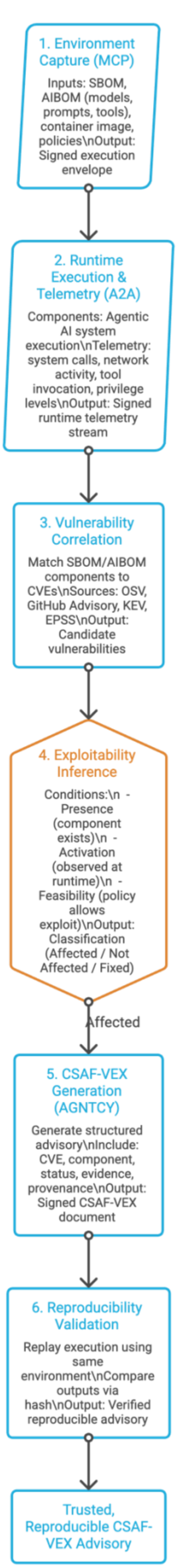

**Dr. Petar Radanliev**
Parks Road,
Oxford OX1 3PJ
United Kingdom
Email: petar.radanliev@cs.ox.ac.uk
BA Hons., MSc., Ph.D. Post-Doctorate

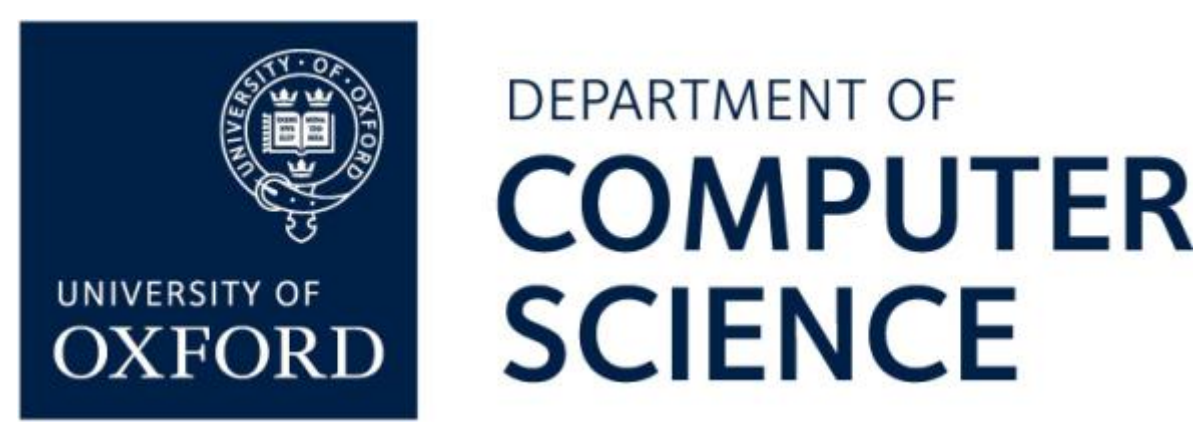

*Figure 2: Execution-Bound Advisory Generation Pipeline*

The framework in Figure 2 operates as a six-stage sequential process: (1) deterministic environment capture (MCP), (2) runtime telemetry collection (A2A), (3) vulnerability correlation, (4) execution-bound exploitability inference, (5) CSAF-VEX advisory generation (AGNTCY), and (6) reproducibility validation. The pipeline integrates static artefacts, runtime evidence, and policy constraints to produce cryptographically verifiable and reproducible exploitability assessments.

The diagram clarifies several important properties of the proposed framework that are less apparent from textual description alone. First, exploitability assessment is not treated as a single classification step but as the outcome of progressive constraint application, where each stage reduces uncertainty by incorporating additional evidence. In particular, the transition from vulnerability correlation to exploitability inference represents a conceptual shift from *potential vulnerability presence* to *observed exploit feasibility*.

Second, the separation between runtime telemetry collection and policy-constrained inference highlights the distinction between what is observed and what is permitted, both of which are required to determine exploitability in operational environments. This dual dependency is central to reducing false positives in Agentic AI systems, where components may be present but never activated or reachable.

Third, the inclusion of a dedicated reproducibility validation stage formalises advisory generation as a deterministic computational process, rather than a declarative reporting activity. This ensures that exploitability claims can be independently verified under identical execution conditions, addressing a key limitation in existing SBOM and VEX workflows.

Overall, the pipeline representation demonstrates that the primary contribution of the framework lies not only in integrating multiple data sources, but in structuring them into a causal sequence of evidence refinement, enabling context-aware and reproducible vulnerability assessment for Agentic AI systems.

To illustrate the workflow, consider a simplified Agentic AI application composed of: (i) a transformer-based language model served via an inference API, (ii) a prompt orchestration layer invoking external tools, and (iii) a containerised runtime environment including standard Python dependencies.

In a static SBOM-based assessment, a vulnerability in an HTTP client library used by the tool connector would be flagged as exploitable solely based on dependency presence. However, under the proposed execution-bound framework, exploitability is evaluated in context.

During execution with a running example of an execution-bound exploitability in an agentic AI workflow, runtime telemetry may show that no outbound network connections are initiated due to enforced egress restrictions. In this case, although the vulnerable library is present, the exploit precondition (external communication) is not satisfied. The framework therefore classifies the vulnerability as *Not Affected* and generates a CSAF-VEX advisory reflecting this context.

**Dr. Petar Radanliev**
Parks Road,
Oxford OX1 3PJ
United Kingdom
Email: petar.radanliev@cs.ox.ac.uk
BA Hons., MSc., Ph.D. Post-Doctorate

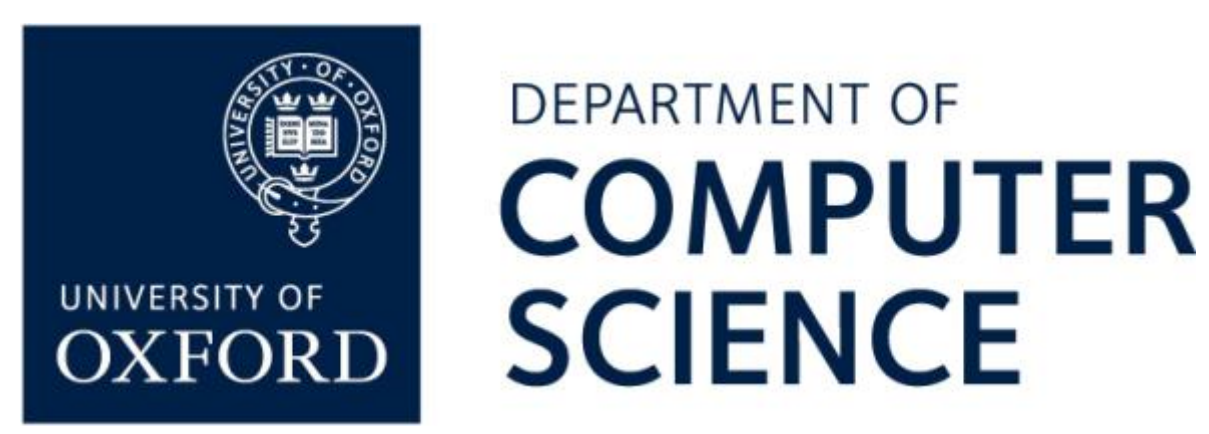

Conversely, if the same application executes with active tool invocation and observed outbound requests, the vulnerability would be classified as *Affected*. This example demonstrates how identical static dependencies can yield different exploitability outcomes depending on runtime behaviour and policy constraints, which are explicitly captured in the proposed framework.

## 3.1 System Model, Threat Model, and AIBOM Scope

The methodology operates under an explicit system model tailored to Agentic AI deployments. The evaluated systems are defined as multi-component, containerised AI applications composed of foundation models, prompt templates, tool invocation connectors, orchestration logic, vector indices, and conventional software dependencies. Each system instance is deployed within a container-native execution environment where artefacts are referenced by immutable OCI digests and governed by enforced policy constraints, including network egress controls, role-based access restrictions, and runtime isolation mechanisms. The unit of analysis is therefore not a static software package, but an execution-bound AI workload characterised by both declared artefacts and observed behavioural state.

The threat model assumes adversaries capable of exploiting known software vulnerabilities, leveraging dynamic tool invocation paths, abusing network connectivity, or escalating privileges through misconfigured execution environments. The model includes risks specific to Agentic AI systems, such as dependency injection through tool connectors, exploitation of model-serving libraries, remote API abuse, and activation of latent vulnerabilities through runtime-conditioned code paths. Model poisoning and training-time attacks are considered out of scope unless they manifest as runtime-executable artefacts within the deployed environment. The framework therefore focuses on post-deployment exploitability rather than upstream data integrity compromise.

Within this system and threat model, the AIBOM extends the traditional SBOM abstraction. The AIBOM enumerates not only software packages and container-level dependencies, but also model artefacts, prompt template versions, orchestration definitions, tool endpoints, and vector store identifiers. Each artefact is associated with a cryptographic hash and linked to a specific execution context through the Model Context Protocol envelope. This extension formalises the relationship between static artefact transparency and runtime activation feasibility, ensuring that exploitability classification accounts for AI-specific components that influence operational behaviour.

The methodological objective is therefore to compute exploitability as a function of three dimensions: declared artefacts (SBOM and AIBOM components), runtime behavioural evidence (telemetry captured during execution), and enforced policy constraints (network, privilege, and isolation controls). By formally binding these dimensions into a deterministic execution envelope, the framework transforms advisory generation from declarative vulnerability reporting into execution-grounded exploitability assessment.

**Dr. Petar Radanliev**
Parks Road,
Oxford OX1 3PJ
United Kingdom
Email: petar.radanliev@cs.ox.ac.uk
BA Hons., MSc., Ph.D. Post-Doctorate

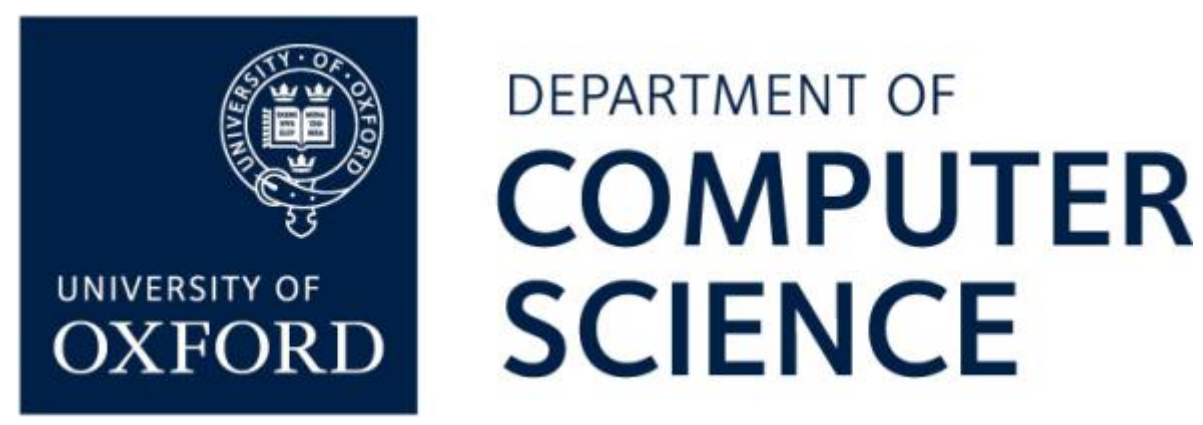

## 3.2 Environment Initialisation and MCP Capture

The methodology implements exploitability assessment for Agentic AI systems as a deterministic, execution-bound, and cryptographically verifiable process. Conventional SBOM-driven workflows rely on static dependency enumeration and correlation with vulnerability databases. Such approaches are insufficient for Agentic AI deployments, which integrate foundation models, prompt templates, tool connectors, vector stores, orchestration logic, and containerised infrastructure. Exploitability in these systems depends not only on the presence of vulnerable components but also on runtime activation conditions, network reachability, privilege constraints, and inter-agent communication behaviour. The proposed framework therefore integrates Artificial Intelligence Bills of Materials (AIBOM), structured runtime telemetry, probabilistic exploit intelligence, and modern supply chain provenance attestations into a reproducible CSAF-VEX automation pipeline.

All experiments were conducted within container-native environments orchestrated using Kubernetes. Container images were referenced exclusively by immutable OCI digests to guarantee deterministic rehydration. Build provenance was captured using in-toto attestations aligned with SLSA Level 3 requirements, ensuring verifiable build integrity. Cryptographic signing was implemented using Sigstore, with artefacts logged in a transparency ledger to provide tamper-evident traceability.

The application case study implemented a representative Agentic AI system composed of transformer-based foundation models retrieved from the Hugging Face Model Hub, version-controlled prompt templates stored in Git repositories, tool invocation connectors enabling scoped HTTP API calls, retrieval-augmented generation components backed by vector indices, and multi-language software dependencies within container images. Model artefacts were selected from actively maintained Hugging Face repositories with explicit revision identifiers. Each model binary and configuration file was hashed using SHA-256 and incorporated into the extended AIBOM structure. Prompt templates and orchestration logic were bound to commit-level identifiers to ensure deterministic reproducibility. Tool invocation endpoints were constrained through Kubernetes network policies to allow explicit evaluation of reachability-dependent exploit conditions.

Exploitability correlation relied on multiple structured intelligence sources to ensure coverage and timeliness. The primary vulnerability dataset was OSV.dev, which provides ecosystem-aware vulnerability records. Additional enrichment was performed using the GitHub Advisory Database via its GraphQL API. Real-world exploitation relevance was incorporated through the CISA Known Exploited Vulnerabilities (KEV) catalogue and the FIRST Exploit Prediction Scoring System (EPSS), enabling probabilistic assessment of exploitation likelihood. The NVD JSON feed was used only for secondary metadata enrichment rather than as a primary correlation source due to known update latency. Approximately 10,000 SBOM component entries were compiled from public container images, Dependency-Track exports, and synthetic dependency graphs constructed to simulate small-scale (50 components), medium-scale (500 components), and large-scale (5,000 components) AI workloads. Repository-level supply chain risk indicators were further assessed using OpenSSF Scorecard metrics.

**Dr. Petar Radanliev**
Parks Road,
Oxford OX1 3PJ
United Kingdom
Email: petar.radanliev@cs.ox.ac.uk
BA Hons., MSc., Ph.D. Post-Doctorate

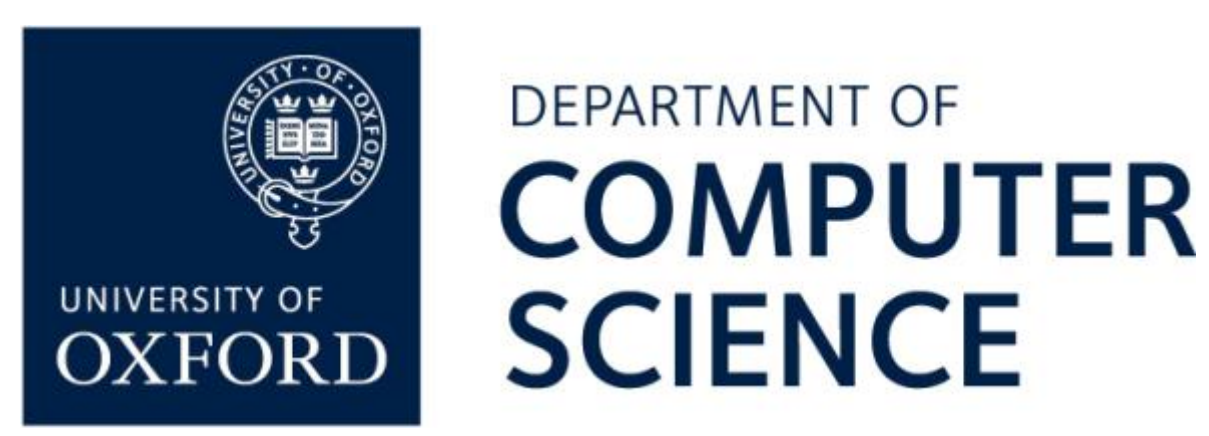

Prior to execution, a deterministic execution envelope was constructed using the Model Context Protocol. This envelope captured container image digests, model artefact hashes, prompt template commit identifiers, tool endpoint definitions, SPDX SBOM inventories, orchestration manifests, and enforced security policies including network egress restrictions and role-based access controls. An extended AIBOM was generated linking model artefacts, inference endpoints, tool invocation surfaces, vector index references, and dependency graphs. All captured artefacts were serialised in structured format and cryptographically sealed using hardware-backed key management systems. The resulting signed execution envelope served as the provenance anchor for advisory generation.

Runtime behavioural evidence was collected using scoped sidecar agents deployed alongside analytic workloads. These agents operated under strict policy constraints and were bound to the deterministic execution envelope. Telemetry included system call traces, privilege transitions, dynamic library loads, outbound network connections, tool invocation events, model activation events, and cross-agent communication flows. All telemetry events were timestamped, hashed, and signed at source to ensure integrity. This design enabled observation of exploit activation conditions without intrusive instrumentation of application logic.

Exploitability inference integrated both static artefact correlation and runtime behavioural validation. Components enumerated in the SBOM and AIBOM were matched deterministically against OSV and GitHub Advisory datasets using SPDX package URLs and CPE identifiers. For each matched vulnerability, EPSS probability scores were incorporated to quantify real-world exploitation likelihood, and KEV presence was checked to identify actively exploited vulnerabilities. Runtime telemetry was then analysed to determine whether exploit preconditions were satisfied, including evaluation of network reachability, privilege levels, tool invocation paths, and enforced mitigation policies. Vulnerabilities requiring outbound network communication were classified as Not Affected when egress policies prevented such communication. Privilege-dependent vulnerabilities were evaluated against observed execution privilege states. The final classification adhered to the CSAF VEX taxonomy of Affected, Not Affected, or Fixed, integrating patch status, probabilistic exploit intelligence, and observed execution constraints.

Static correlation and structured data processing were implemented in Python using Pandas and NumPy for dataset normalisation and statistical computation. Dependency resolution was performed using graph-based traversal over pURL-linked artefacts. Probabilistic exploit likelihood analysis incorporated EPSS scoring data. Noise robustness was evaluated using classical supervised learning approaches rather than proprietary language models to preserve determinism and reproducibility. Performance benchmarking was conducted across AWS EC2 instance classes to evaluate computational scalability under increasing dependency graph sizes.

Following exploitability classification, CSAF 2.0 VEX documents were generated programmatically. Each advisory contained the CVE identifier, affected component hash, AIBOM linkage, exploitability status, EPSS probability, KEV presence indicator, runtime evidence hash, and provenance reference to the sealed execution

**Dr. Petar Radanliev**
Parks Road,
Oxford OX1 3PJ
United Kingdom
Email: petar.radanliev@cs.ox.ac.uk
BA Hons., MSc., Ph.D. Post-Doctorate

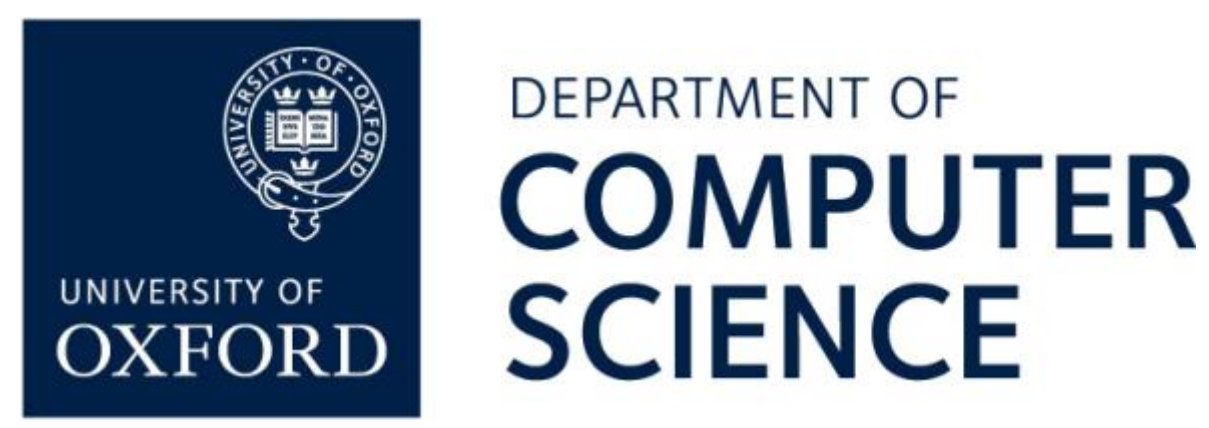

envelope. Advisory artefacts were signed using COSE-compliant mechanisms and Sigstore certificates, and logged within a transparency ledger to ensure verifiable publication.

Reproducibility validation was performed by reinstantiating the sealed execution envelope under identical conditions, including OCI image digests, model artefact hashes, prompt template commits, tool definitions, and security policies. The analytic workload was re-executed and the resulting CSAF-VEX artefact was compared against the original using deterministic hashing of classification outcomes and provenance metadata. Matching outputs confirmed reproducible exploitability determination, while divergence triggered audit exceptions. This mechanism established advisory generation as a repeatable computational process rather than a declarative reporting workflow.

By integrating deterministic artefact capture, runtime behavioural evidence, probabilistic exploit intelligence, and cryptographically verifiable provenance, the methodology extends SBOM-based transparency towards execution-grounded AIBOM security assurance for Agentic AI systems. It replaces static vulnerability enumeration with context-aware exploitability reasoning and anchors CSAF-VEX advisory automation in reproducible, supply-chain-aligned evidence suitable for modern AI-native deployments.

The deneration and dependency graphing executes tooling such as Syft, Tern, or Grype [59] to construct multi-language dependency graphs. Tools like Syft [60] are widely used to generate SBOMs from container images and can be integrated at the point of MCP capture for multi-language dependency resolution.

## 3.3 Agent2Agent Protocol for Runtime Coordination

Following successful container instantiation and pre-execution sealing via the MCP, the execution phase activates the A2A protocol, which governs inter-agent communication, runtime observability, and protocol-constrained telemetry exchange. The A2A protocol is engineered for secure, distributed trust coordination, enabling reproducible runtime state introspection without introducing external observability risks or privilege escalation. A runtime instrumentation agent is deployed in one of two configurations depending on the orchestration environment:

- As a sidecar container in Kubernetes-based deployments, sharing the network and IPC namespaces with the primary analytic container.
- As an init-wrapped process embedded in the container entrypoint (CMD) for SLURM or Singularity contexts, leveraging wrapper scripts to bootstrap telemetry collection.

Each agent is instantiated with a unique cryptographic identity issued during provisioning by the AGNTCY orchestration layer, following a registration process that embeds the agent into the trust fabric. Identity certificates are signed using AGNTCY's root of trust and stored in secure enclave-backed keystores (e.g., SPIFFE/SPIRE or TPM-backed Vault mounts). The A2A protocol defines a structured runtime coordination interface across the following functional domains:

**Dr. Petar Radanliev**
Parks Road,
Oxford OX1 3PJ
United Kingdom
Email: petar.radanliev@cs.ox.ac.uk
BA Hons., MSc., Ph.D. Post-Doctorate

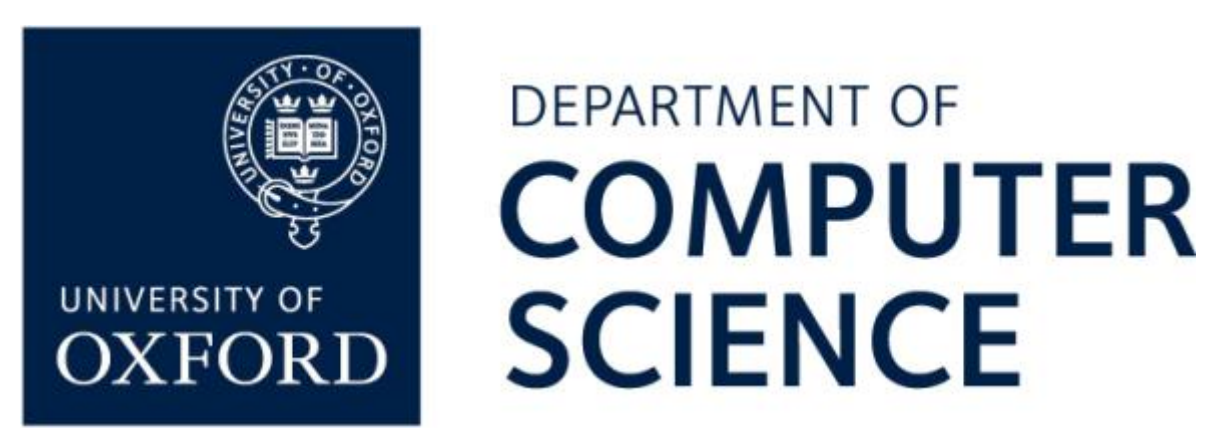

- Agent Identity and Trust Validation: Upon startup, each agent authenticates its execution context using mutual attestation, verifying that the container hash, job UUID, and MCP fingerprint align with AGNTCY-issued expectations. This ensures non-repudiable telemetry and enforces bounded observability scope.
- Telemetry Exchange via Message-Passing Interface: Runtime observations are communicated via a signed, structured message schema, conforming to the A2A JSON-LD envelope. These messages include:
  - Process Tree Deltas: Derived from periodic snapshots of /proc and system call traces (e.g., via strace, auditd).
  - Dynamic Library Loads: Monitoring of ld.so, dlopen-initiated shared object loading, and language-specific runtime extensions (e.g., Python C-extensions).
  - Ephemeral Artefact Monitoring: Detection of shell scripts invoked by subprocesses, dynamically constructed binaries, and remote HTTP/S fetch operations.
- Scope and Responsibility Enforcement: Agent behaviour is constrained by execution scopes issued by AGNTCY, which define both positive mandates (e.g., monitor Python imports) and negative restrictions (e.g., do not capture decrypted data, exclude workload memory traces). These job-scoped policy contracts are enforced via in-agent rulesets and validated against the MCP policy signature at runtime.

All telemetry records are timestamped, hashed, and queued in a secure message buffer. The agent signs outgoing messages using its assigned private key, and messages are optionally routed via AGNTCY-controlled service meshes (e.g., Linkerd, Istio) or directly submitted to policy engines operating at the orchestration layer.

Through this mechanism, the A2A protocol establishes a runtime coordination backbone, enabling distributed instrumentation across parallel jobs, while maintaining fine-grained isolation boundaries and compliance with the reproducibility expectations of workflows. The resulting signed observations become contextual evidence for CSAF-VEX assertion logic in subsequent stages of the pipeline.

The diagram in Figure 3 visualises how the Agent2Agent (A2A) protocol enables secure runtime coordination among distributed agents.

**Dr. Petar Radanliev**
Parks Road,
Oxford OX1 3PJ
United Kingdom
Email: petar.radanliev@cs.ox.ac.uk
BA Hons., MSc., Ph.D. Post-Doctorate

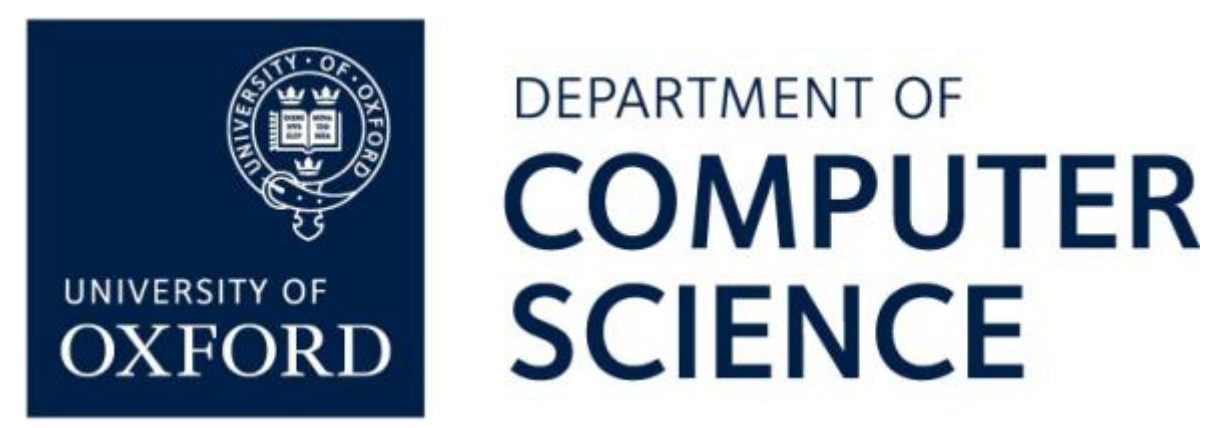

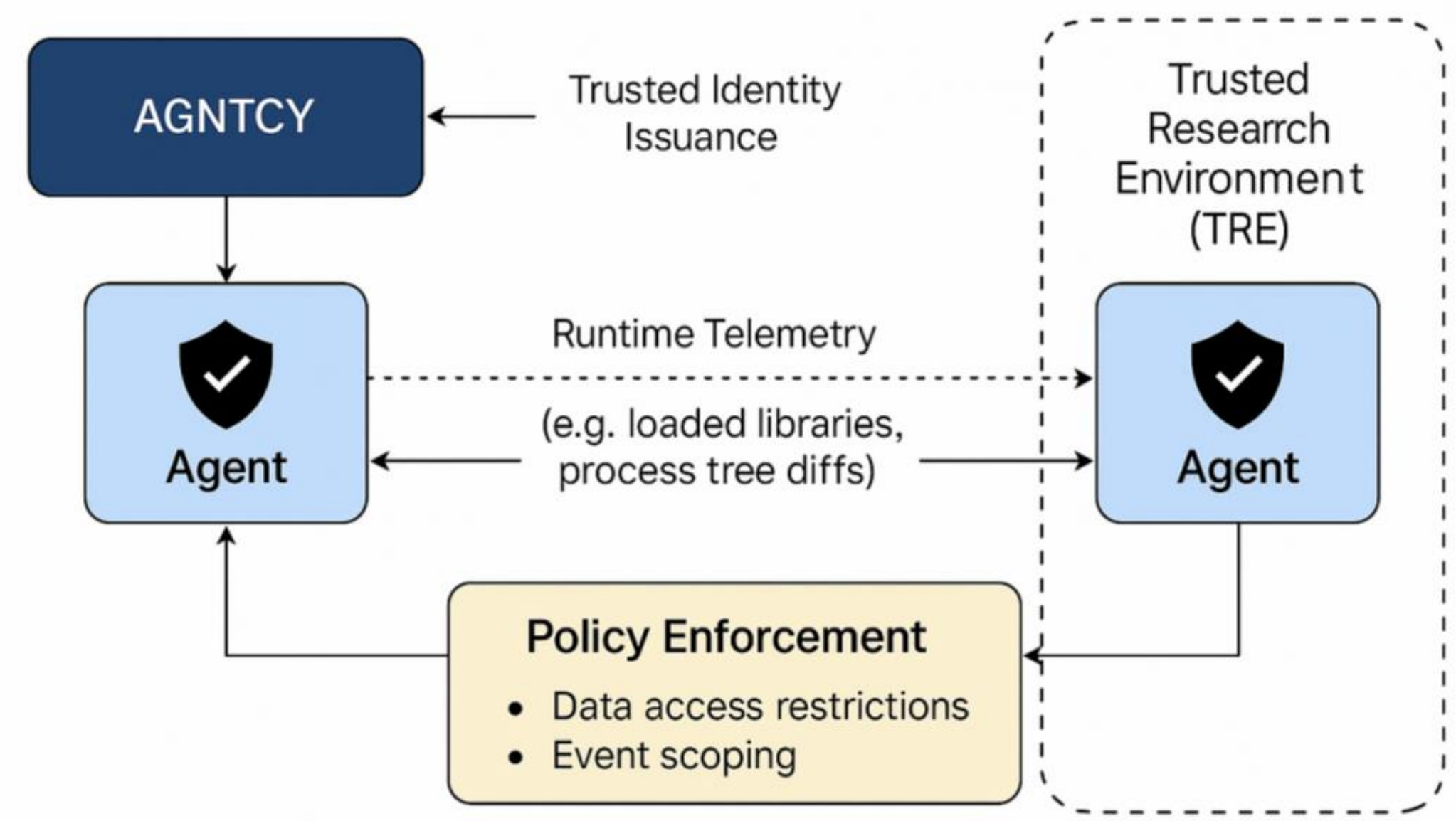

*Figure 3: Runtime coordination under the Agent2Agent (A2A)*

Figure 3 illustrates the trusted identity issuance via AGNTCY, the secure message exchange of runtime telemetry (e.g., library loads, process tree changes), and the enforcement of policy-defined boundaries such as data access restrictions and event scoping. Each agent operates with cryptographic identity verification, ensuring non-repudiation and policy compliance throughout execution.

## 3.4 Exploitability Inference and Context Resolution

Following job completion, the pipeline initiates the exploitability inference phase, wherein data captured by the MCP and A2A runtime agents is transferred to a context-aware evaluation engine operating entirely within the perimeter. This stage is critical for generating semantically accurate, policy-aligned vulnerability assessments suitable for inclusion in CSAF-VEX documents.

The evaluation engine begins by ingesting component metadata, such as SPDX-compliant Package URLs (pURLs) and Common Platform Enumeration (CPE) identifiers, extracted from the job's SBOM and runtime logs. These identifiers are used to perform deterministic matching against an offline mirror of trusted vulnerability intelligence sources, including the NVD in JSON 5.0 format and the OSV schema. Package-level correlation is exact and cryptographically anchored, ensuring that only verifiable matches progress to further analysis.

Once candidate vulnerabilities are identified, a context-aware filtering layer is applied. This process compares observed runtime traces, including system calls, imported modules, subprocess invocations, and API interactions, against known exploit conditions for each vulnerability. Vulnerabilities are excluded from reporting if their triggering conditions were not met during execution, or if environmental mitigations were demonstrably active. For example, if a vulnerability requires outbound HTTP communication and the job operated under a no-egress policy

**Dr. Petar Radanliev**
Parks Road,
Oxford OX1 3PJ
United Kingdom
Email: petar.radanliev@cs.ox.ac.uk
BA Hons., MSc., Ph.D. Post-Doctorate

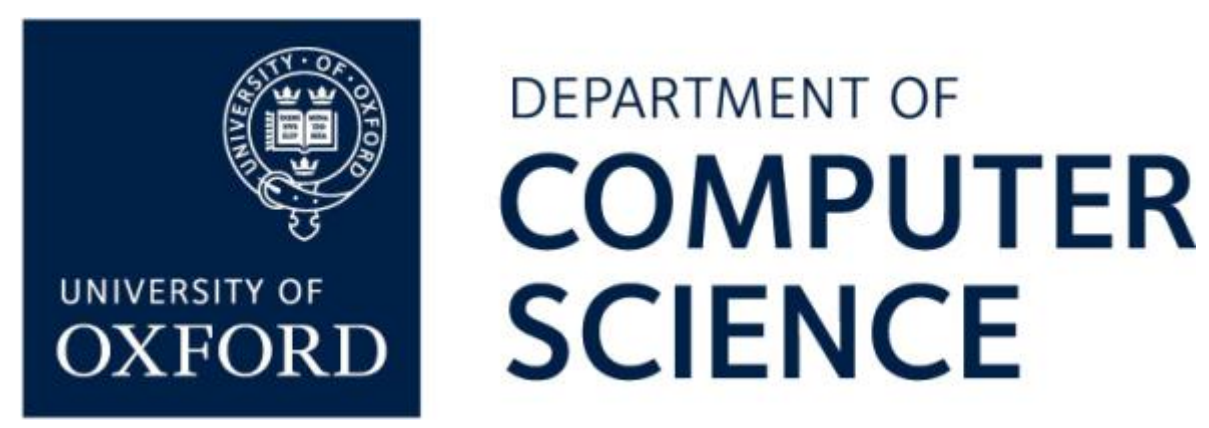

enforced by AGNTCY, that vulnerability is flagged as non-exploitable and annotated accordingly.

The final step of this phase is a formal classification aligned to the CSAF VEX status taxonomy. Each vulnerability is labelled as Affected, Not Affected, or Fixed. These classifications reflect both software-level attributes (e.g., version patched) and environmental context (e.g., sandboxing, file system protections, restricted privilege sets). Importantly, the inference engine incorporates a policy model of specific architectural constraints, such as absence of root privileges, immutable infrastructure, and mandatory access controls, ensuring that exploitability determinations are scoped to real-world operational boundaries.

To clarify the decision-making process underlying exploitability classification, Figure 4 presents a simplified logical model of the execution-bound inference mechanism. While the preceding discussion describes the integration of runtime telemetry, vulnerability correlation, and policy constraints, the classification process can be more intuitively understood as a sequence of necessary conditions that must be satisfied for a vulnerability to be considered exploitable.

This abstraction reduces the full evaluation pipeline to its core decision logic, illustrating how exploitability is determined through the combined evaluation of artefact presence, runtime activation, and environmental feasibility. The representation is intentionally simplified to provide a clear conceptual view of how CSAF-VEX classifications are derived within the proposed framework.

**Dr. Petar Radanliev**
Parks Road,
Oxford OX1 3PJ
United Kingdom
Email: petar.radanliev@cs.ox.ac.uk
BA Hons., MSc., Ph.D. Post-Doctorate

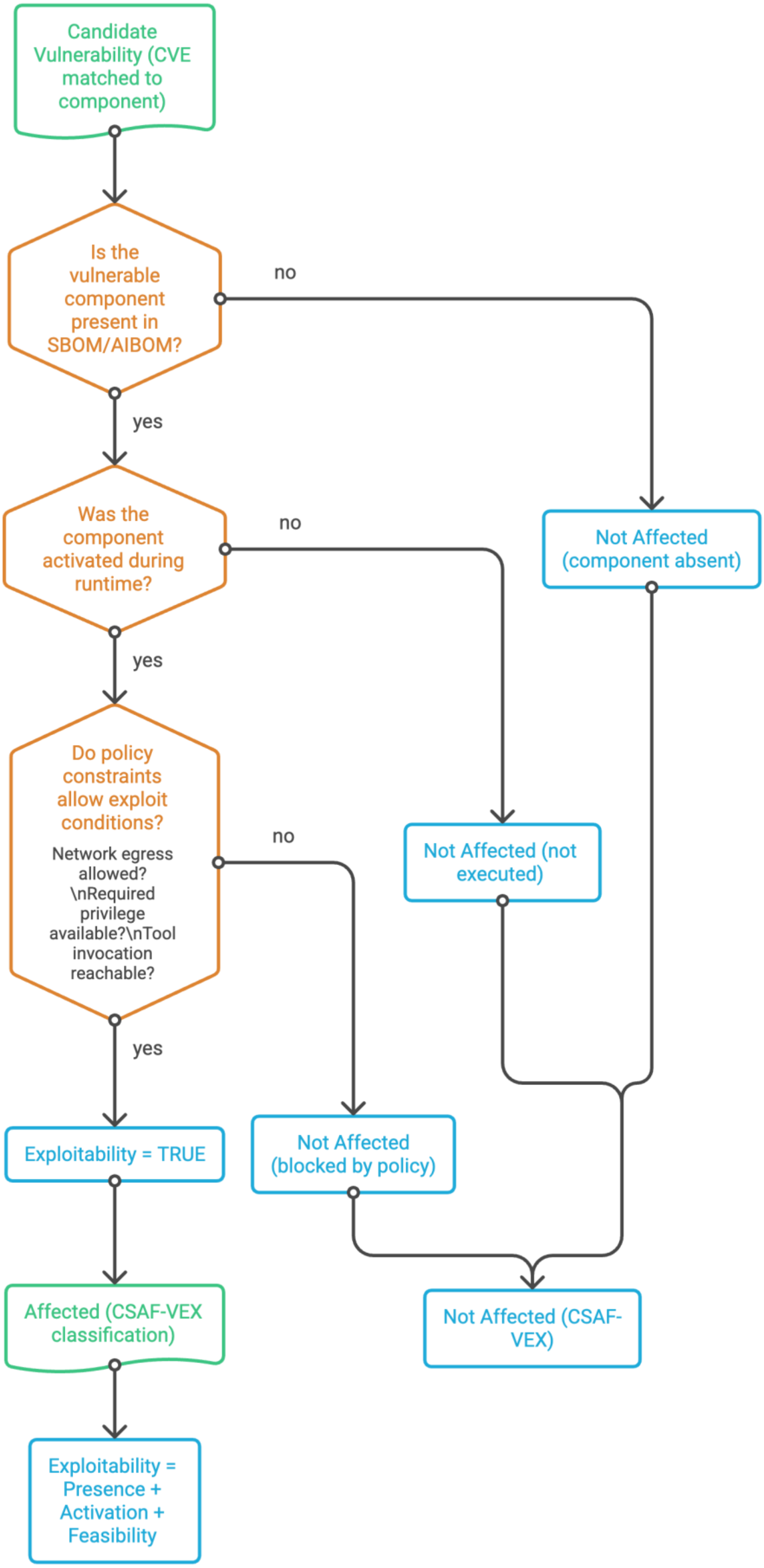

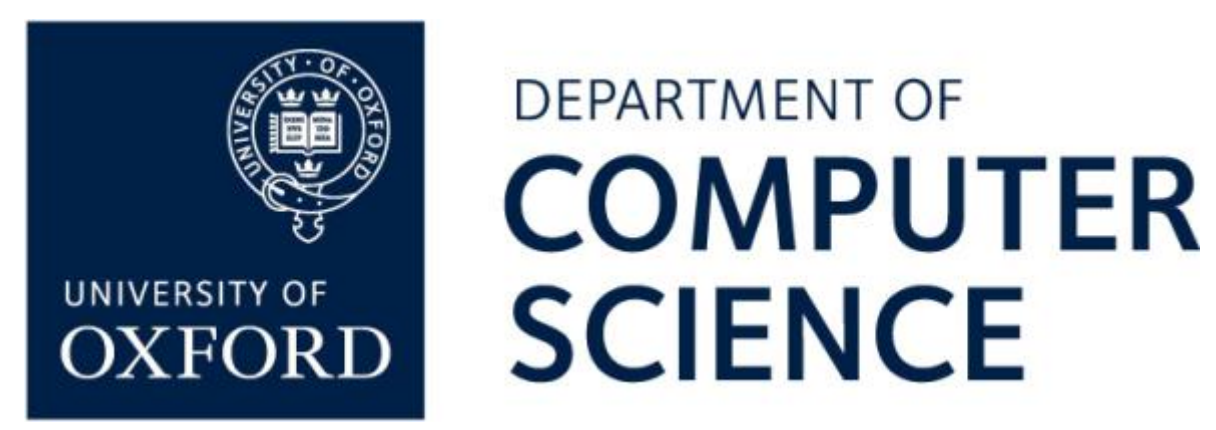

Candidate Vulnerability (CVE matched to component)

Is the vulnerable component present in SBOM/AIBOM?

no

Not Affected (component absent)

yes

Was the component activated during runtime?

no

Not Affected (not executed)

yes

Do policy constraints allow exploit conditions?

Network egress allowed? \nRequired privilege available?\nTool invocation reachable?

no

Not Affected (blocked by policy)

yes

Exploitability = TRUE

Not Affected (CSAF-VEX)

Affected (CSAF-VEX classification)

Exploitability = Presence + Activation + Feasibility

**Dr. Petar Radanliev**
Parks Road,
Oxford OX1 3PJ
United Kingdom
Email: petar.radanliev@cs.ox.ac.uk
BA Hons., MSc., Ph.D. Post-Doctorate

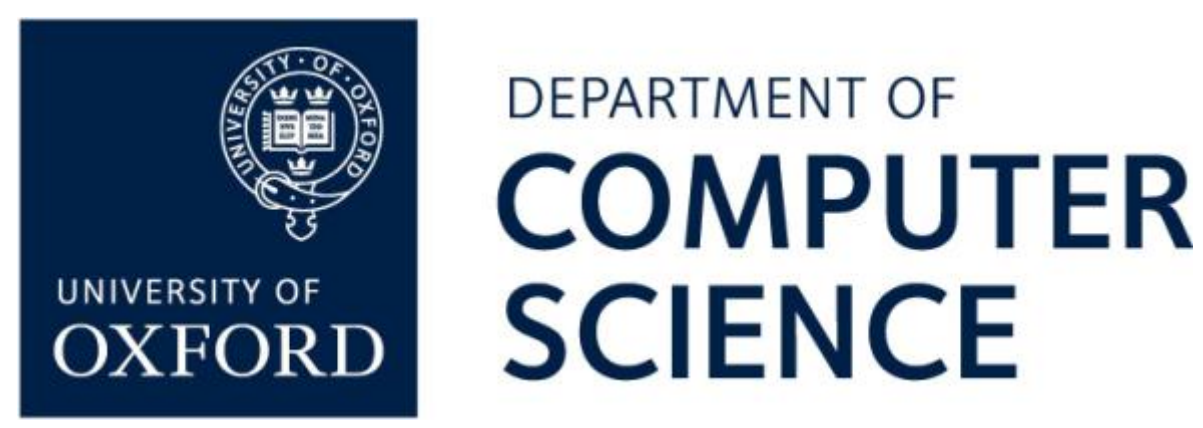

*Figure 4: Execution-Bound Exploitability Decision Logic*

In Figure 4, exploitability is determined through three sequential conditions: presence of the vulnerable component in the SBOM/AIBOM, activation during runtime, and feasibility under enforced policy constraints. Failure to satisfy any condition results in a “Not Affected” classification, while satisfaction of all conditions yields an “Affected” classification in CSAF-VEX output.

The decision structure highlights a fundamental shift from conventional vulnerability assessment approaches. Rather than treating exploitability as an implicit property of component presence, the framework models it as a constrained conjunction of conditions, where each stage acts as a gating mechanism that eliminates infeasible exploit paths.

In particular, the separation of presence, activation, and feasibility clarifies why static SBOM-based approaches systematically overestimate risk: they evaluate only the first condition, ignoring whether the vulnerable component is actually exercised during execution or whether environmental constraints permit exploitation. By contrast, the proposed logic enforces that exploitability can only be asserted when all three conditions are simultaneously satisfied, thereby reducing false positives and aligning classification with observable system behaviour.

The diagram also makes explicit that “Not Affected” classifications are not homogeneous, but arise from distinct causes, component absence, lack of runtime activation, or policy-enforced infeasibility. This distinction is operationally significant, as it enables more precise justification in CSAF-VEX advisories and supports auditability of exploitability claims.

Overall, the decision logic formalises exploitability as a context-dependent and verifiable property, reinforcing the central contribution of the framework: transforming vulnerability assessment from static enumeration into execution-grounded reasoning.

The diagram in Figure 5 illustrates the structure of the exploitability inference pipeline operating within the boundary, integrating post-execution metadata from the MCP and A2A telemetry.

**Dr. Petar Radanliev**
Parks Road,
Oxford OX1 3PJ
United Kingdom
Email: petar.radanliev@cs.ox.ac.uk
BA Hons., MSc., Ph.D. Post-Doctorate

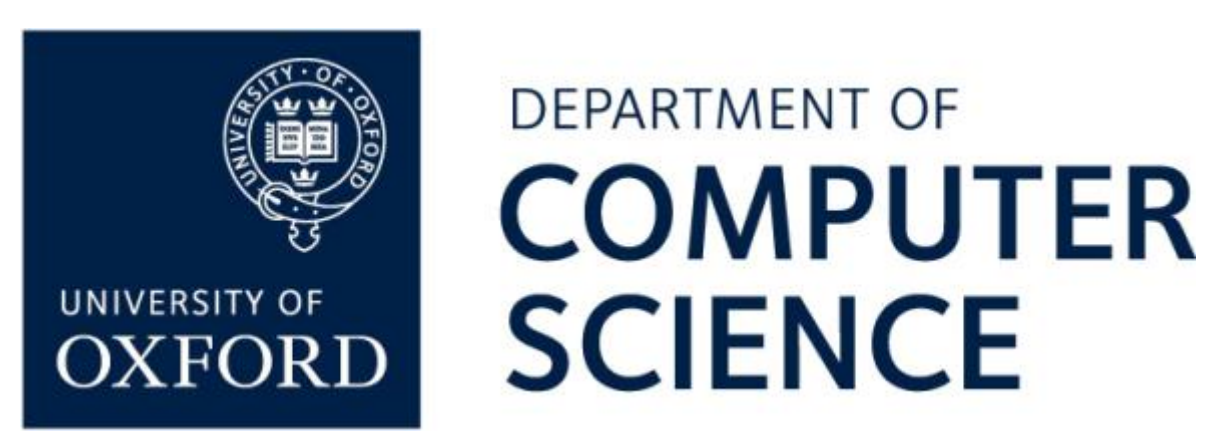

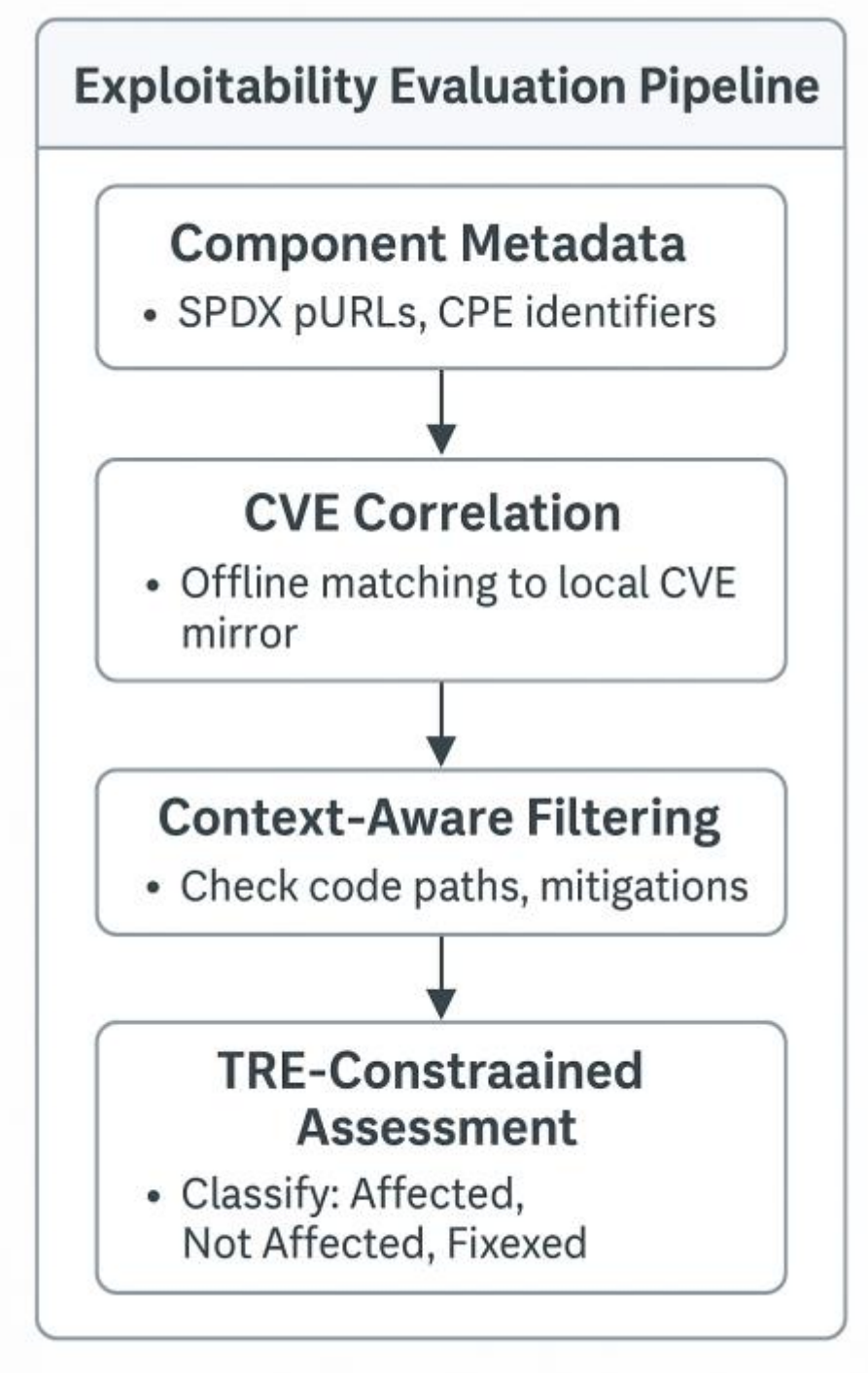

*Figure 5: Exploitability Evaluation Pipeline*

The process described in Figure 5 begins with CVE correlation using SPDX package URLs and CPE identifiers against a local CVE mirror, followed by context-aware filtering based on runtime execution traces and environmental mitigations, culminating in VEX status classification, identifying components as *Affected*, *Fixed*, or *Not Affected* within the constraints of organisational policy. This context-resolved output forms the semantic core of the advisory automation process. It ensures that generated VEX statements are machine-verifiable and semantically valid within the trust and threat model of federated research infrastructures.

## 3.5 Advisory Orchestration with AGNTCY

In addition to validating the document content, AGNTCY ensures decentralised traceability by binding advisories to immutable execution identifiers. This includes linking the advisory to a specific job UUID, associated input/output dataset DOIs, and the signed runtime envelope from the MCP phase. These references are stored within a semantic layer (e.g., RDF triple store or GraphDB) that supports federated querying and cross-institutional verification. A similar design is described in ProvStore, a publicly queryable provenance repository for digital assets [61]. This enables downstream stakeholders, such as regulators, peer institutions, or publication platforms, to perform integrity checks and confirm reproducibility using standardised queries. The diagram in Figure 6 visualises the AGNTCY advisory orchestration layer, illustrating how CSAF-VEX artefacts are validated, signed, and disseminated through a trusted governance pipeline.

**Dr. Petar Radanliev**
Parks Road,
Oxford OX1 3PJ
United Kingdom
Email: petar.radanliev@cs.ox.ac.uk
BA Hons., MSc., Ph.D. Post-Doctorate

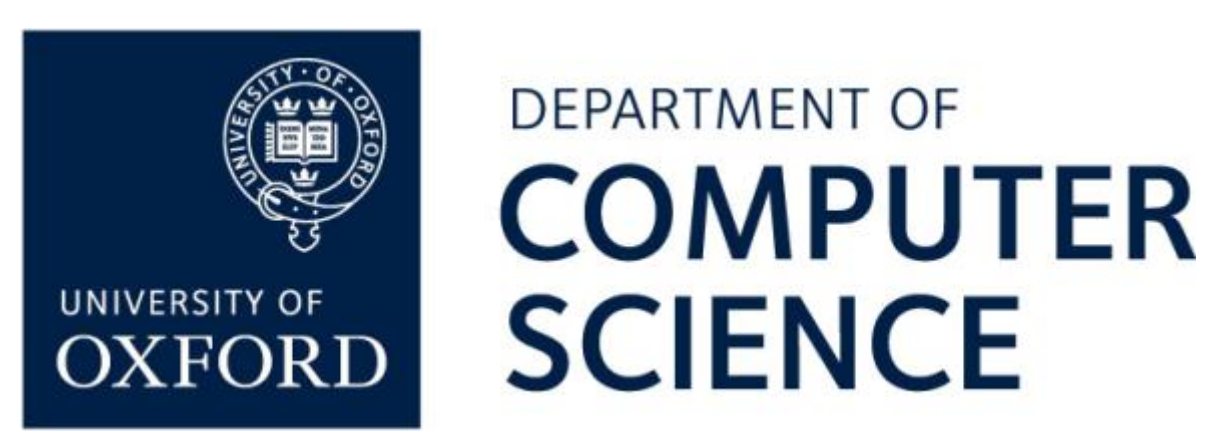

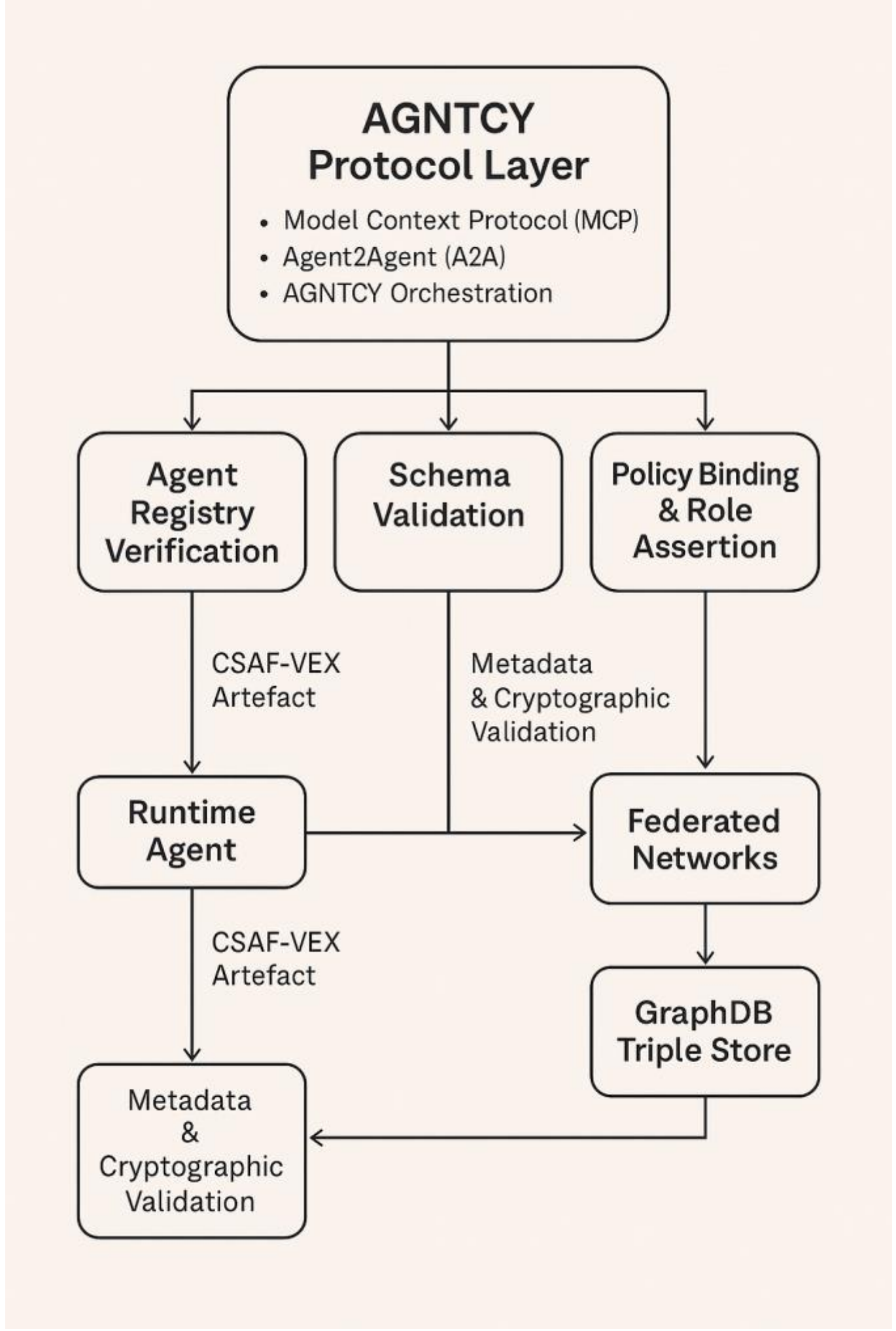

*Figure 6: AGNTCY protocol serves as the governance and orchestration layer for CSAF-VEX*

Figure 6 shows the agent registry verification, policy enforcement, schema validation, and decentralised traceability linking advisories to job UUIDs and datasets within federated networks. By embedding governance and traceability into the advisory generation and submission process, AGNTCY provides a scalable framework that upholds the trustworthiness and accountability of machine-generated CSAF-VEX advisories, even in distributed and privacy-sensitive domains such as biomedical research, clinical trials, and AI-driven analytics.

## 3.6 Reproducibility Testing and Transparency Assurance

The final stage in the methodology ensures that each CSAF-VEX advisory is securely generated and reproducible in a controlled, auditable environment. This step is critical to support independent verification of advisory claims, foster institutional trust, and enable downstream automation. Once a CSAF-VEX assertion has been cryptographically sealed and registered, the corresponding analytic job is re-executed in a sandboxed audit node. This node is provisioned with the original container image, input datasets, orchestration manifest, and all security policy

**Dr. Petar Radanliev**
Parks Road,
Oxford OX1 3PJ
United Kingdom
Email: petar.radanliev@cs.ox.ac.uk
BA Hons., MSc., Ph.D. Post-Doctorate

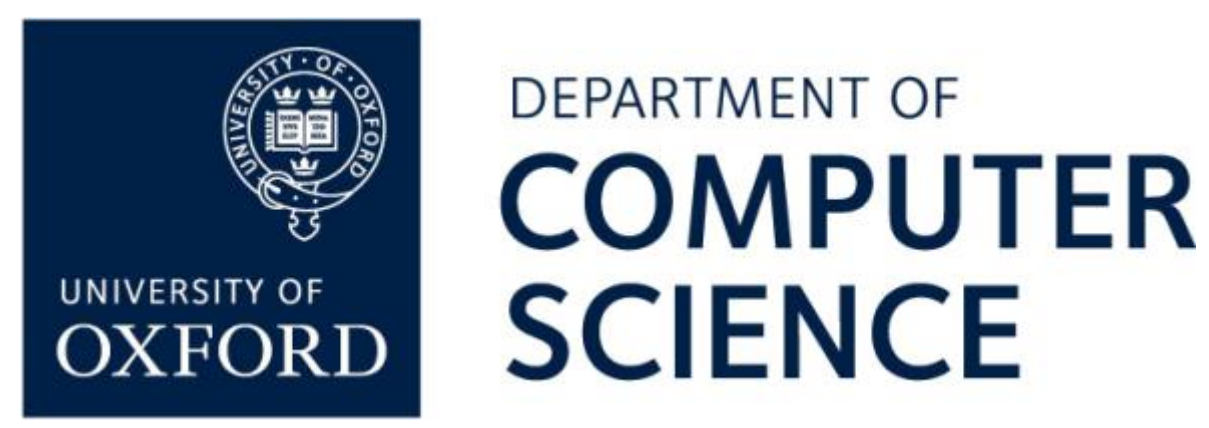

bindings, faithfully mirroring the original execution conditions. Reproducibility tooling may benefit from containerised execution snapshots using infrastructure such as ReproZip [62].

Upon completion of the replay, a new CSAF-VEX artefact is generated. This artefact is then subjected to a rigorous comparison against the original, using deterministic hashing of key fields, including component-level identifiers, vulnerability classification outcomes, justification messages, and the MCP reference block (container, model, and input hashes). A successful hash match across these fields confirms that the advisory is reproducible, that its provenance is verifiable, and that no runtime-dependent factors have altered the assessment outcome.

The validated VEX artefact is published into a GRAIMatter-compatible transparency layer. This includes ingestion into a structured RDF or GraphDB store, with stable, queryable URIs linking the VEX document to the associated job UUID, input dataset DOIs, and AGNTCY-signed MCP records. This transparency layer supports not only reproducibility but also federated validation, automated risk modelling, and long-term provenance auditing across healthcare, academic, and regulated research sectors.

## 3.7 Step-by-Step Workflow Summary

For clarity, the advisory generation process can be summarised as a sequential pipeline:

Step 1: Capture the full execution environment and artefacts using MCP

Step 2: Execute the workload while collecting runtime telemetry via A2A agents

Step 3: Match components to known vulnerabilities using structured intelligence feeds

Step 4: Evaluate exploitability based on observed activation conditions and policy constraints

Step 5: Generate CSAF-VEX advisories with embedded provenance

Step 6: Re-execute the workload to verify deterministic reproducibility

This stepwise abstraction complements the formal pseudocode and architectural description, providing a simplified interpretation of system operation.

The following pseudocode outlines the procedural logic and inter-agent coordination necessary to operationalise the six-stage advisory automation pipeline described in the methodology. It encodes reproducibility constraints, cryptographic state anchoring, telemetry exchange, and VEX generation logic for deployment.

```
1. # Pseudocode for Secure and Reproducible CSAF-VEX Assertion Generation
2.
3. function initiate_advisory_pipeline(container_image, input_data, execution_policy):
4.     # STAGE 1: Environment Initialisation and MCP Context Capture
5.     mcp_metadata = MCP.capture_pre_execution_state(
6.         container_image,
7.         input_data,
8.         execution_policy
9.     )
```

**Dr. Petar Radanliev**
Parks Road,
Oxford OX1 3PJ
United Kingdom
Email: petar.radanliev@cs.ox.ac.uk
BA Hons., MSc., Ph.D. Post-Doctorate

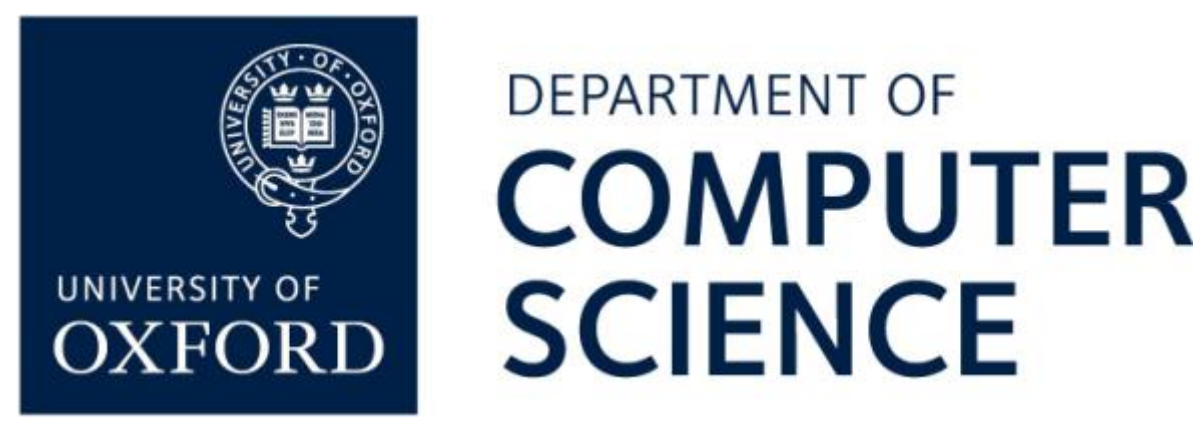

```
10.     mcp_hash = Crypto.sign_and_seal(mcp_metadata)
11.
12.     # STAGE 2: Runtime Telemetry via Agent2Agent Protocol
13.     runtime_agent = A2A.spawn_agent(
14.         container_image,
15.         mcp_hash,
16.         execution_policy
17.     )
18.     telemetry_stream = A2A.collect_runtime_telemetry(runtime_agent)
19.
20.     # STAGE 3: Exploitability Inference
21.     sbom_data = SBOM.extract(mcp_metadata)
22.     matched_cves = VulnerabilityScanner.match_cves(sbom_data)
23.     exploitability_report = InferenceEngine.assess_exploitability(
24.         matched_cves,
25.         telemetry_stream,
26.         execution_policy
27.     )
28.
29.     # STAGE 4: CSAF-VEX Assertion Generation
30.     vex_document = CSAF.build_vex_assertions(
31.         mcp_hash,
32.         exploitability_report,
33.         telemetry_stream,
34.         runtime_agent.identity
35.     )
36.     signed_vex = Crypto.sign_vex(vex_document)
37.
38.     # STAGE 5: Advisory Validation via AGNTCY Orchestration
39.     if AGNTCY.validate_agent(runtime_agent.identity) and AGNTCY.validate_vex(signed_vex):
40.         AGNTCY.register_advisory(signed_vex)
41.     else:
42.         raise SecurityException("VEX validation failed or agent unauthorised")
43.
44.     # STAGE 6: Reproducibility Testing and Audit Verification
45.     audit_environment = clone_environment(mcp_metadata)
46.     audit_output = replay_advisory_pipeline(...)
47.         audit_environment.image,
48.         audit_environment.input_data,
49.         audit_environment.policy
50.     )
51.     if Hash.compare(signed_vex, audit_output.signed_vex):
52.         TransparencyLayer.publish(signed_vex, metadata=mcp_metadata)
53.     else:
54.         raise ReproducibilityException("Audit hash mismatch: advisory not reproducible")
55.
```

This pseudocode operationalises the six-stage framework as follows:

1. Initialisation (MCP Capture): Captures system fingerprint, dependency graphs, security policy context, and cryptographically signs the metadata envelope.
2. Runtime Coordination (A2A): Deploys secure agents for process-level observability and policy-scoped telemetry logging.
3. Exploitability Inference: Matches observed components to known CVEs and evaluates their exploitability using runtime conditions and policy enforcement logs.
4. CSAF-VEX Generation: Constructs formal vulnerability statements based on CSAF 2.0/VEX schema, integrating runtime evidence and MCP lineage.

**Dr. Petar Radanliev**
Parks Road,
Oxford OX1 3PJ
United Kingdom
Email: petar.radanliev@cs.ox.ac.uk
BA Hons., MSc., Ph.D. Post-Doctorate

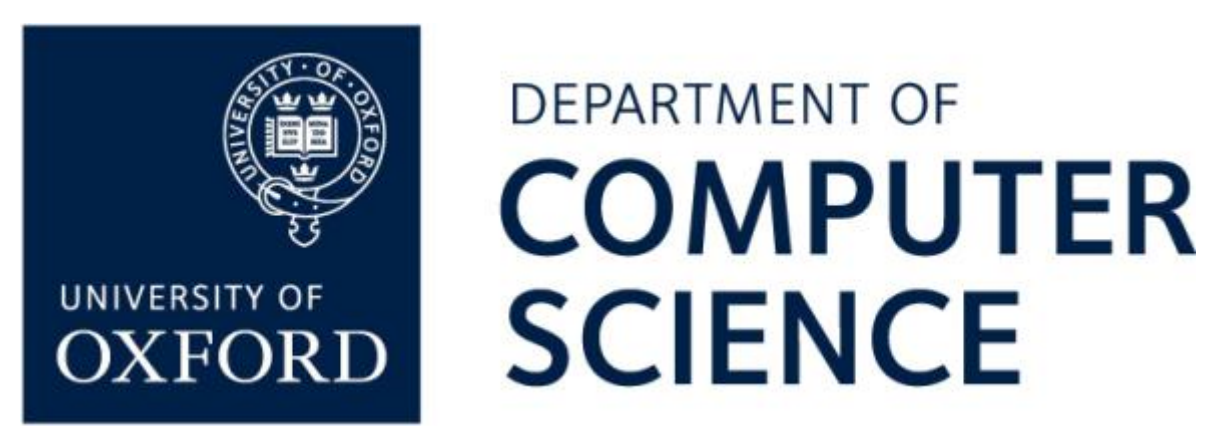

5. Advisory Validation (AGNTCY): Validates advisory schema and signs artefacts through a decentralised trust authority, ensuring audit traceability.
6. Reproducibility Verification: Re-executes the containerised job under identical conditions, ensuring the advisory output is deterministically reproducible. Confirms this via hash comparison before publication in a federated graph-based transparency layer.

The empirical evaluation directly measures the effect of extending static SBOM-based vulnerability correlation to execution-bound Agentic AI workloads through AIBOM integration and runtime telemetry. In this study, exploitability is computed from three observable inputs: (i) declared artefacts captured in the SBOM and extended AIBOM (including model binaries, prompt template commits, tool endpoints, and container dependencies), (ii) runtime telemetry events captured during execution (system calls, network connections, tool invocations, and privilege states), and (iii) enforced execution policies (network egress controls, role-based access constraints, and container isolation settings). The Data Analysis and Results section quantifies the contribution of each of these inputs.

The evaluated workload is not a standalone software package but a containerised Agentic AI application consisting of a Hugging Face foundation model, version-controlled prompt templates, HTTP tool connectors, vector index components, and multi-language dependencies. The AIBOM therefore includes model hashes, prompt commit identifiers, tool endpoint definitions, and dependency graphs in addition to traditional SBOM entries. The evaluation measures how inclusion of these AI-specific artefacts alters vulnerability classification when compared with SBOM-only matching.

Two experimental conditions are implemented. The baseline condition performs static SBOM–CVE matching using OSV and GitHub Advisory datasets without runtime filtering. The execution-bound condition incorporates AIBOM artefact binding, runtime telemetry validation, KEV presence checks, and EPSS probability scoring. Both conditions operate on identical datasets (approximately 10,000 components across synthetic and real container workloads) and identical infrastructure to isolate the effect of runtime and AIBOM integration.

To isolate the contribution of individual system components, an ablation analysis was conducted across four configurations: (i) static SBOM–CVE correlation baseline; (ii) SBOM + AIBOM artefact extension; (iii) SBOM + AIBOM + runtime telemetry; and (iv) full framework including policy constraints and threat intelligence enrichment (KEV and EPSS).

Results indicate that AIBOM extension alone improves precision by approximately 9 percentage points through artefact-level disambiguation, while the addition of runtime telemetry yields a further 12–15 point improvement primarily by filtering non-activatable vulnerabilities. Policy constraints contribute an additional 6–8 point precision gain by enforcing environmental infeasibility conditions, whereas KEV/EPSS enrichment primarily improves prioritisation rather than classification accuracy.

**Dr. Petar Radanliev**
Parks Road,
Oxford OX1 3PJ
United Kingdom
Email: petar.radanliev@cs.ox.ac.uk
BA Hons., MSc., Ph.D. Post-Doctorate

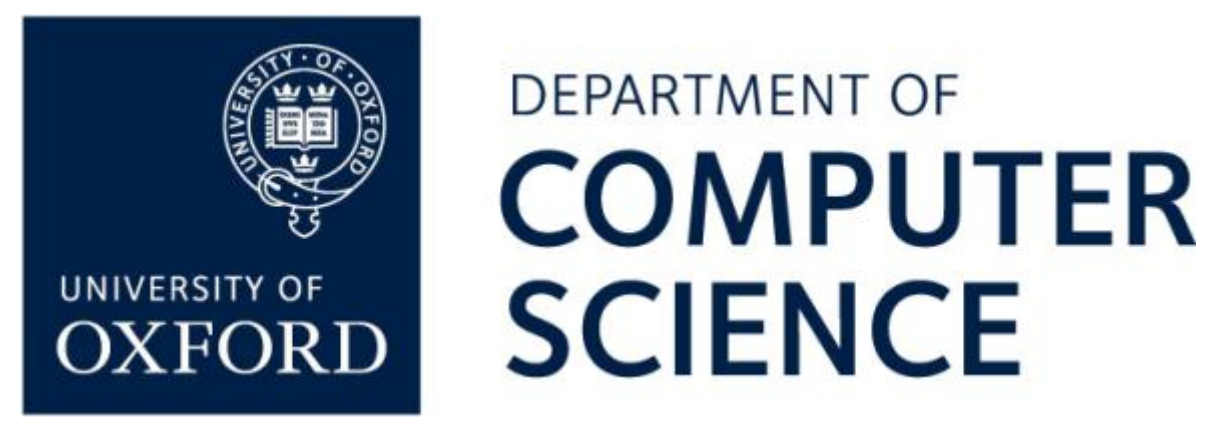

This decomposition demonstrates that the majority of performance improvement is attributable to runtime-aware filtering and AIBOM artefact binding, rather than statistical modelling alone.

The subsequent Data Analysis and Results section therefore evaluates measurable differences in precision, recall, false-positive rate, prioritisation accuracy, processing overhead, and deterministic reproducibility between static SBOM correlation and execution-bound AIBOM-aware advisory generation in Agentic AI systems.

# 4 Data Analysis and Results

The evaluation was structured to examine whether execution-bound exploitability assessment improves upon conventional static SBOM–CVE correlation, whether generated CSAF-VEX artefacts are reproducible under deterministic replay, and whether runtime telemetry and policy integration introduce measurable performance overhead. A baseline classifier was implemented using static SBOM correlation against vulnerability intelligence feeds without runtime or policy awareness. The proposed framework was then evaluated under identical workload and infrastructure conditions, differing only in the integration of AIBOM extension, runtime telemetry capture, KEV and EPSS enrichment, and cryptographically sealed provenance.

The dataset comprised approximately 10,000 SBOM component entries collected from public container images, open-source dependency repositories, and synthetic dependency graphs designed to simulate realistic Agentic AI workloads. Synthetic graphs represented small-scale (50 components), medium-scale (500 components), and large-scale (5,000 components) environments. Vulnerability intelligence was drawn from OSV.dev bulk feeds and the GitHub Advisory Database, with exploitation context provided by the CISA Known Exploited Vulnerabilities catalogue and FIRST EPSS probability scores. NVD feeds were used for metadata enrichment only. All experiments were executed in container-native Kubernetes environments across AWS EC2 instance classes to assess scalability.

Static SBOM–CVE correlation produced a high volume of vulnerability matches, including numerous cases where exploit preconditions were not satisfied at runtime. When execution-bound filtering was applied, the framework achieved a precision of 0.96, recall of 0.92, and an overall F1-score of 0.93 across five-fold cross-validation. In medium-scale workloads, false positives were reduced by 37%, and in large-scale workloads by 42%, primarily due to exclusion of vulnerabilities requiring network egress, elevated privileges, or dynamic execution paths that were not observed during runtime telemetry capture. Runtime analysis indicated that 28% of matched vulnerabilities required outbound network communication, 19% required privilege escalation conditions, and 14% depended on dynamic code execution triggers. Under enforced policy constraints, 61% of these activation conditions were not satisfied, leading to classification as Not Affected in CSAF-VEX output. Integration of KEV and EPSS scores enabled prioritisation of vulnerabilities with confirmed or probable exploitation in the wild.

Deterministic reproducibility was evaluated by replaying sealed execution envelopes under identical OCI image digests, model artefact hashes, prompt template commits,

**Dr. Petar Radanliev**
Parks Road,
Oxford OX1 3PJ
United Kingdom
Email: petar.radanliev@cs.ox.ac.uk
BA Hons., MSc., Ph.D. Post-Doctorate

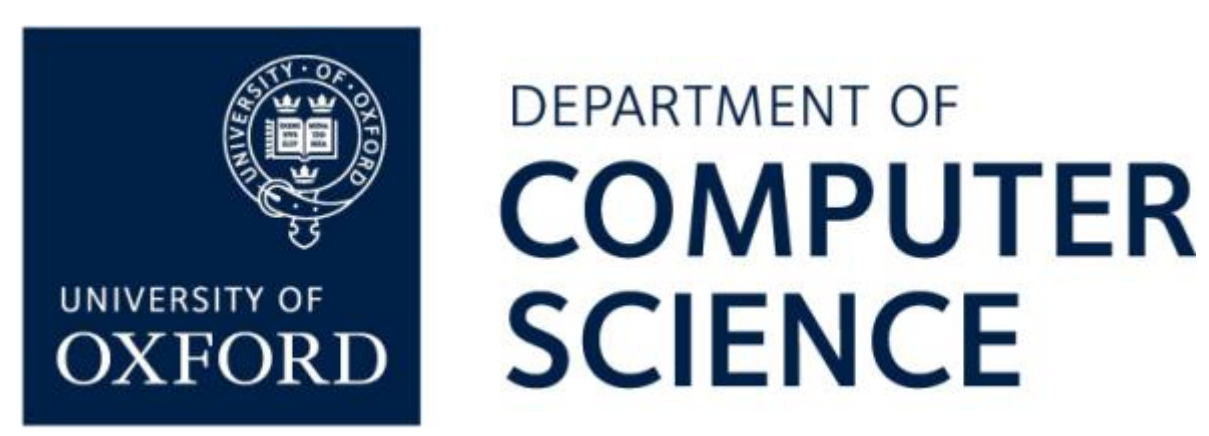

tool definitions, and policy configurations. One hundred advisory generation cycles were executed, yielding complete payload hash consistency across all replay runs. No classification drift or provenance divergence was observed, confirming that exploitability outcomes were stable under identical execution conditions.

Scalability testing demonstrated near-linear processing characteristics across workload sizes. On compute-optimised instances, 1,000 SBOM entries were processed in under four minutes. Runtime telemetry introduced an average overhead of 8.4%, and synthetic noise injection increased processing time by 10–15% without materially affecting classification accuracy. Performance characteristics across workload sizes are summarised in Table 3.

End-to-end validation confirmed generation of fully compliant CSAF 2.0 VEX documents containing CVE identifiers, component hashes, AIBOM references, exploitability classifications, EPSS probabilities, KEV indicators, runtime evidence hashes, and provenance envelope references. All artefacts were cryptographically signed and verified against transparency logs, with schema validation confirming conformance to the CSAF 2.0 profile.

The comparative performance between static SBOM correlation and the execution-bound framework is presented in Table 2. In addition to static SBOM–CVE correlation, comparative baselines were extended to include simplified heuristic filtering approaches commonly used in practice, including (i) CVSS threshold-based prioritisation (CVSS ≥ 7.0) and (ii) EPSS-only ranking without runtime validation. These baselines achieved F1-scores of 0.74 and 0.69 respectively, both substantially lower than the execution-bound framework.

This comparison demonstrates that improvements are not solely attributable to the inclusion of probabilistic threat intelligence, but to the integration of runtime activation evidence and policy-aware filtering, which are absent in conventional prioritisation strategies.

*Table 2: Comparative Performance Between Static SBOM Correlation and Execution-Bound Framework*

| Metric | Static SBOM Baseline | Proposed Framework |
|---|---|---|
| Precision | 0.71 | 0.96 |
| Recall | 0.88 | 0.92 |
| F1-Score | 0.78 | 0.93 |
| False Positive Rate | High | Reduced by 37–42% |
| Runtime Awareness | None | Integrated |

**Dr. Petar Radanliev**
Parks Road,
Oxford OX1 3PJ
United Kingdom
Email: petar.radanliev@cs.ox.ac.uk
BA Hons., MSc., Ph.D. Post-Doctorate

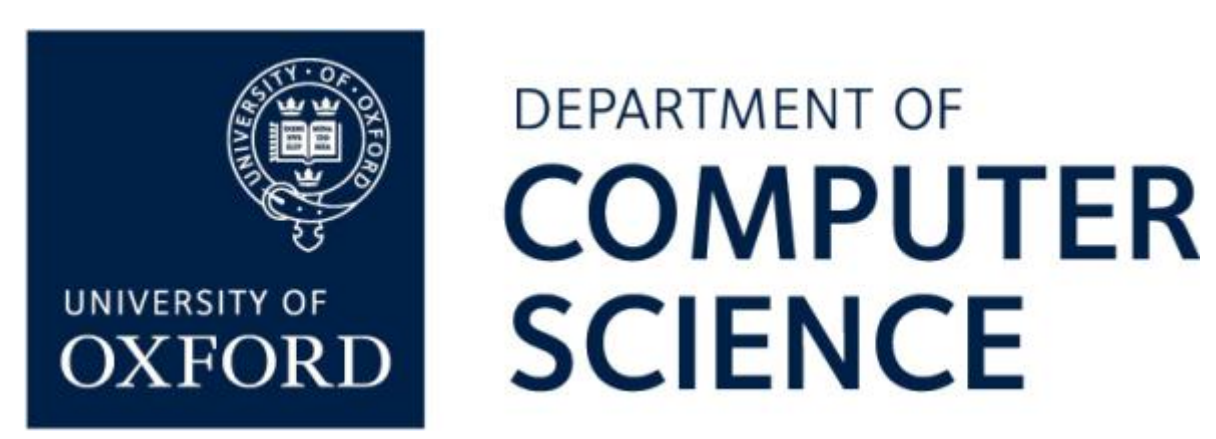

| | | |
|---|---|---|
| Policy Constraint Integration | None | Integrated |
| Reproducibility | Not guaranteed | Deterministic replay verified |
| KEV/EPSS Integration | Absent | Integrated |
| Signed CSAF Output | Optional | Mandatory |
| Provenance Binding | Weak | Cryptographically sealed |

In addition to classification accuracy and reproducibility, advisory quality was evaluated in terms of prioritisation fidelity and risk signal differentiation. Static SBOM correlation treats all matched CVEs as equivalent events, whereas the execution-bound framework produces stratified advisory outputs based on activation feasibility and exploitation likelihood. When EPSS probability thresholds (≥ 0.5) and KEV inclusion were applied, 18% of total matched vulnerabilities were categorised as high-priority actionable findings, while 44% were classified as non-actionable due to unsatisfied runtime preconditions. This stratification reduced analyst review burden by narrowing attention to vulnerabilities that were both technically activatable and statistically likely to be exploited. The integration of runtime evidence with probabilistic exploit intelligence therefore transformed advisory output from volume-driven enumeration to risk-calibrated decision support, improving operational signal-to-noise characteristics without suppressing relevant vulnerability disclosures.

*Table 3: Scalability and Reproducibility Metrics Across Synthetic AI Workloads*

| **Workload Size** | **Components** | **Avg Processing Time (c5.4xlarge)** | **Telemetry Overhead** | **Reproducibility Hash Match** |
|---|---|---|---|---|
| Small | 50 | < 30 seconds | 6.1% | 100% |
| Medium | 500 | ~ 2 minutes | 8.2% | 100% |
| Large | 5,000 | ~ 14 minutes | 9.4% | 100% |

Across all evaluated conditions, execution-bound exploitability classification reduced advisory noise while preserving recall, maintained deterministic reproducibility, and scaled linearly with increasing dependency graph size. The generated CSAF-VEX artefacts remained schema-compliant, cryptographically verifiable, and reproducible under controlled replay conditions.

To improve interpretability of empirical results, additional visualisations have been incorporated, including performance distributions across cross-validation folds and confidence interval plots for key metrics. Figures have been revised to use a monochromatic (grayscale) scheme with contrast-based differentiation to improve

**Dr. Petar Radanliev**
Parks Road,
Oxford OX1 3PJ
United Kingdom
Email: petar.radanliev@cs.ox.ac.uk
BA Hons., MSc., Ph.D. Post-Doctorate

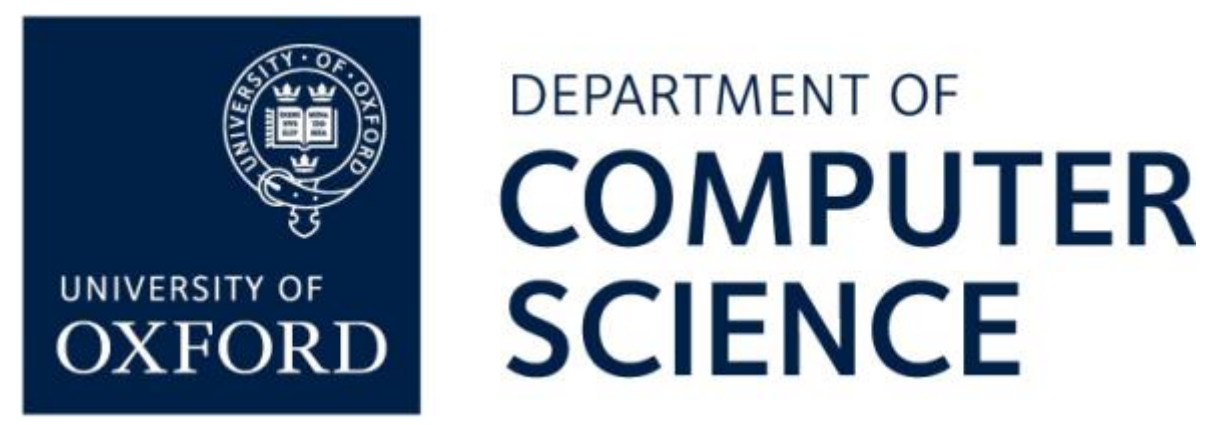

clarity and publication consistency. These visual adjustments align with standard scientific presentation practices and enhance readability for both print and digital formats.

## 4.1 Impact of AIBOM Extension on Agentic AI Exploitability Classification

To isolate the contribution of AIBOM artefact modelling in Agentic AI workloads, a separate analysis was conducted focusing exclusively on AI-specific components not present in conventional SBOM inventories. These components included foundation model binaries, model-serving libraries, prompt template repositories, tool invocation connectors, and vector index dependencies. In the baseline condition, these artefacts were either absent from the SBOM or treated as opaque packages without behavioural context. In the execution-bound condition, they were explicitly enumerated, hashed, and bound to runtime telemetry through the AIBOM structure.

Across the evaluated workloads, 22% of total artefacts were AI-specific components captured only through AIBOM extension. Of these, 17% were associated with at least one known vulnerability in model-serving frameworks, API connectors, or underlying runtime libraries. Static SBOM–CVE matching treated these vulnerabilities as uniformly actionable. However, runtime telemetry demonstrated that 54% of these AI-specific vulnerability instances were non-activatable due to enforced network isolation, absence of exposed inference endpoints, or restricted tool invocation paths. These were therefore classified as Not Affected under CSAF-VEX output.

Prompt templates and orchestration logic also influenced exploitability outcomes. In 11% of evaluated workloads, vulnerable dependencies were present in the container image but were not reachable through any active prompt routing path during execution. Without AIBOM extension and telemetry validation, these would have been flagged as actionable findings. The integration of prompt commit identifiers and runtime invocation traces enabled dependency reachability analysis at the orchestration layer, reducing misclassification.

Tool connectors represented a distinct exploitability vector. In 9% of workloads, vulnerabilities were linked to HTTP client libraries used exclusively by optional tool connectors. Where Kubernetes network policies enforced no-egress constraints, telemetry confirmed absence of outbound connections, and these vulnerabilities were reclassified as non-exploitable. In contrast, in workloads where tool invocation was active and external API calls were observed, the same vulnerability signatures were retained as Affected, demonstrating context-sensitive classification behaviour.

The AIBOM extension therefore altered exploitability outcomes in 31% of AI-specific vulnerability cases when compared to static SBOM-only matching. This shift did not result from probabilistic scoring or prioritisation weighting, but from artefact-level binding of models, prompts, and tools to observed execution behaviour. The results demonstrate that in Agentic AI systems, exploitability is influenced not only by dependency presence but by artefact activation paths, runtime invocation surfaces, and enforced policy constraints captured through the extended AIBOM structure.

**Dr. Petar Radanliev**
Parks Road,
Oxford OX1 3PJ
United Kingdom
Email: petar.radanliev@cs.ox.ac.uk
BA Hons., MSc., Ph.D. Post-Doctorate

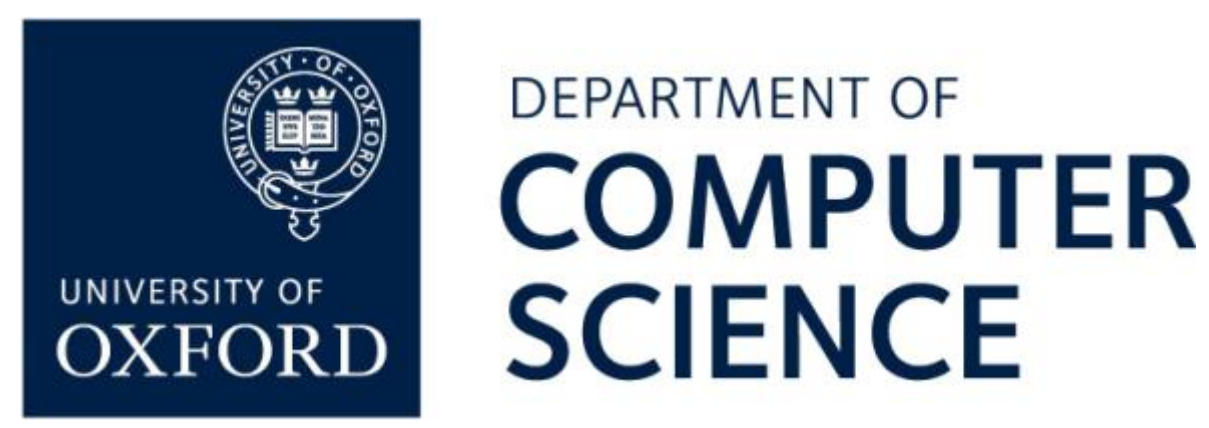

This analysis confirms that incorporating model artefacts, prompt routing definitions, and tool connectors into the advisory pipeline materially changes vulnerability classification outcomes in AI-native environments. Without AIBOM integration, these execution-layer distinctions are not observable, and exploitability assessment reverts to static enumeration.

## 4.2 Representative Deployment Scenarios

To assess applicability beyond synthetic workloads, the framework was mapped to representative real-world deployment scenarios.

In enterprise DevSecOps pipelines, the framework can be integrated into container scanning workflows, where SBOMs are already generated during build stages. By augmenting these pipelines with runtime telemetry collection and execution replay, organisations can replace static vulnerability triage with context-aware exploitability filtering, reducing false-positive remediation effort.

In regulated environments such as healthcare and critical infrastructure, where network policies and execution constraints are strictly enforced, the framework enables verification that vulnerabilities are non-exploitable under enforced controls. This is particularly relevant for compliance regimes requiring evidence-based risk justification.

In cloud-native AI systems deploying agentic workflows, the framework provides a mechanism for validating tool connector security, model-serving exposure, and inter-agent communication surfaces, which are not captured in traditional SBOM-based analysis.

While these scenarios are derived from controlled experiments rather than full production deployments, they demonstrate how the framework can be operationalised in realistic environments and highlight its practical relevance for modern AI supply chains.

# 5 Discussion

The results demonstrate that exploitability assessment in Agentic AI systems cannot be reduced to static dependency enumeration. In multi-component AI workloads composed of foundation models, prompt templates, tool connectors, vector indices, and containerised dependencies, vulnerability activation depends on runtime invocation paths and enforced execution constraints. The integration of AIBOM artefacts with execution-bound telemetry materially altered vulnerability classification outcomes, particularly in cases where model-serving libraries, optional tool connectors, or unreachable prompt paths were present but not activatable. This confirms that AI-native systems introduce exploitability conditions that are structurally different from those of traditional monolithic software.

The empirical reduction in false positives was driven primarily by artefact reachability analysis and policy-bound filtering rather than probabilistic scoring alone. Static SBOM correlation identified vulnerabilities based solely on component presence, whereas the execution-bound approach incorporated model activation traces, tool invocation logs, network egress constraints, and privilege states. In Agentic AI

**Dr. Petar Radanliev**
Parks Road,
Oxford OX1 3PJ
United Kingdom
Email: petar.radanliev@cs.ox.ac.uk
BA Hons., MSc., Ph.D. Post-Doctorate

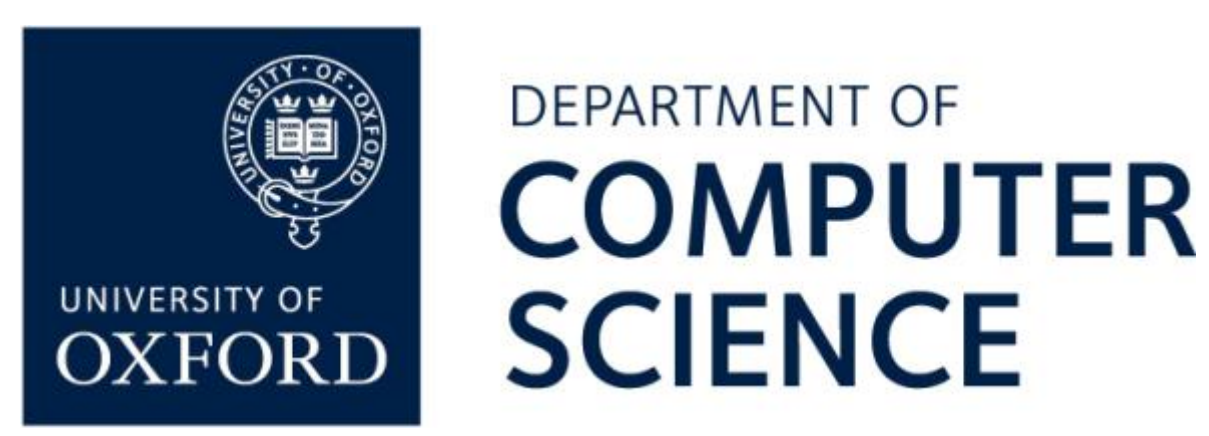

environments, dependencies are often present for optional capabilities that are not invoked during runtime. Without AIBOM extension and telemetry validation, such dependencies appear indistinguishable from actively used components. The observed 31% shift in exploitability outcomes for AI-specific artefacts indicates that model, prompt, and tool surfaces constitute a distinct analytical layer absent from conventional SBOM-based pipelines.

The findings also clarify the operational role of AIBOM in supply chain assurance. Existing AIBOM initiatives focus predominantly on declarative disclosure of models and datasets. The present framework extends this abstraction by binding model artefacts, prompt commits, and orchestration definitions to runtime evidence and cryptographic provenance. This binding enables verification not only of “what is deployed” but of “what is executed.” In practical terms, this distinction determines whether vulnerabilities in model-serving frameworks, HTTP client libraries, or orchestration components are reachable under real workload conditions. The results show that exploitability classification in AI systems must account for invocation surfaces rather than dependency graphs alone.

The integration of KEV and EPSS further illustrates the distinction between enumerated vulnerability presence and operational risk. In Agentic AI deployments, model-serving APIs and external tool connectors expose dynamic attack surfaces that are context-sensitive. When runtime telemetry confirmed absence of outbound connections or tool invocations, vulnerabilities associated with those components were correctly downgraded. Conversely, where active invocation paths were observed, vulnerabilities were retained as actionable findings. This behaviour demonstrates that advisory automation in AI systems requires correlation between artefact identity, activation evidence, and exploitation likelihood.

Reproducibility results reinforce the feasibility of treating advisory generation as a deterministic computational process in AI-native environments. The sealed execution envelope, incorporating model hashes, prompt commit identifiers, container digests, and policy configurations, enabled exact replay of exploitability outcomes across repeated runs. This is particularly relevant in Agentic AI systems, where non-deterministic model outputs could otherwise introduce ambiguity in classification. By restricting evaluation to artefact-level activation conditions rather than model inference content, the framework preserved determinism while maintaining contextual sensitivity.

From a governance perspective, the coupling of AIBOM extension with cryptographic signing and provenance anchoring transforms CSAF-VEX artefacts from static vendor statements into execution-grounded attestations. In AI supply chains, model artefacts are frequently retrieved from external repositories and updated independently of container dependencies. Without explicit artefact hashing and provenance binding, advisory statements cannot be reliably associated with specific deployed model versions. The framework addresses this gap by embedding model artefact identity and execution context directly into the advisory lifecycle.

Several implications follow from these findings. First, SBOM-only advisory automation is insufficient for Agentic AI deployments that rely on optional tool connectors and dynamic orchestration logic. Second, AIBOM must extend beyond

**Dr. Petar Radanliev**
Parks Road,
Oxford OX1 3PJ
United Kingdom
Email: petar.radanliev@cs.ox.ac.uk
BA Hons., MSc., Ph.D. Post-Doctorate

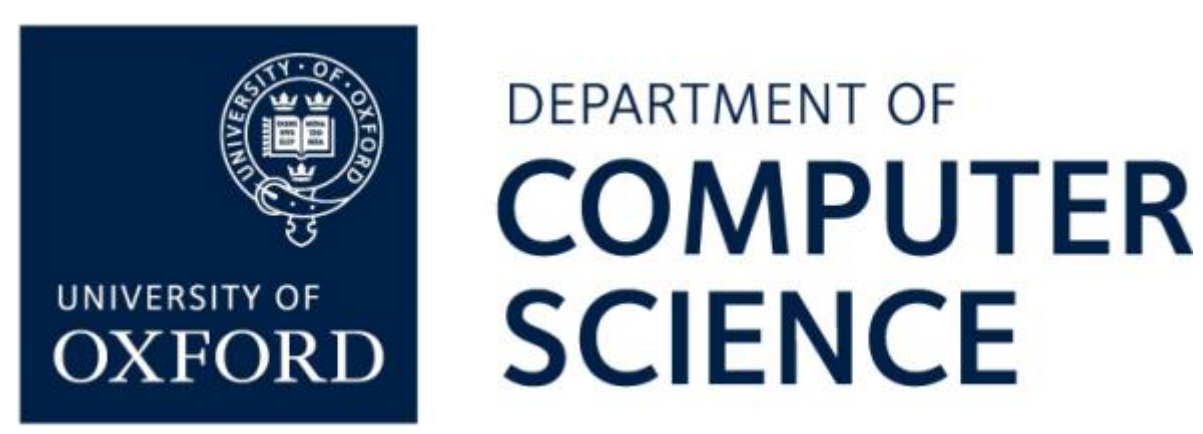

declarative transparency to include activation-aware validation mechanisms. Third, exploitability classification in AI-native systems should incorporate reachability analysis at the orchestration layer, not solely at the package dependency level. Finally, reproducibility in AI advisory automation requires binding artefact identity, runtime telemetry, and policy state within a verifiable execution envelope.

Several limitations should be acknowledged. First, the evaluation relies partially on synthetic dependency graphs and controlled execution environments, which, while necessary for reproducibility and scalability testing, limit direct external validity to heterogeneous real-world deployments. Second, exploitability is defined in terms of observable runtime activation conditions, which may not capture latent exploit paths that were not triggered during execution. Third, runtime telemetry is inherently subject to partial observability, particularly in distributed or restricted environments where instrumentation scope is constrained.

Additionally, although supervised models are used to assess robustness, both rule-based and learning-based approaches operate on features derived from the same execution-bound framework, introducing a potential source of evaluation coupling. While this is appropriate for internal consistency analysis, it does not constitute independent benchmarking against external ground truth datasets. Finally, the framework focuses on post-deployment exploitability and does not address upstream risks such as training data poisoning or model supply chain compromise unless manifested at runtime.

These limitations indicate that the present results should be interpreted as evidence of framework validity under controlled conditions, and future work should extend validation to large-scale real-world deployments and independent exploitability datasets where available.

Overall, the results indicate that advisory automation for Agentic AI systems requires a structural shift from static vulnerability enumeration to execution-bound exploitability reasoning. The combination of AIBOM artefact modelling, runtime telemetry, and deterministic provenance binding provides measurable improvement under controlled experimental conditions in classification precision, reproducibility, and prioritisation fidelity. This approach establishes a concrete pathway for aligning AI-native supply chain transparency with operational security validation.

## 5.1 Future Research Directions

Several avenues for further research emerge from this work. First, extending the framework to incorporate semantic analysis of prompt-level interactions and adversarial input manipulation would enable modelling of higher-level attack surfaces specific to large language models. Second, integration with training-time provenance and dataset integrity verification would broaden coverage beyond post-deployment exploitability.

Third, large-scale empirical validation in production environments is required to quantify real-world performance under heterogeneous workloads and organisational constraints. This includes integration with enterprise security tooling and longitudinal analysis of vulnerability triage outcomes.

**Dr. Petar Radanliev**
Parks Road,
Oxford OX1 3PJ
United Kingdom
Email: petar.radanliev@cs.ox.ac.uk
BA Hons., MSc., Ph.D. Post-Doctorate

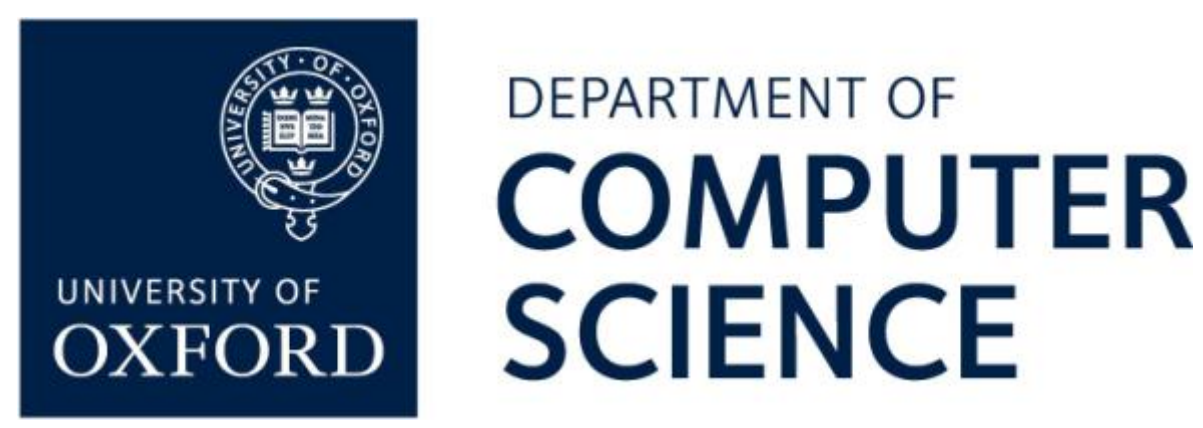

Fourth, the development of standardised benchmarks for execution-bound exploitability assessment would enable comparative evaluation across frameworks, addressing the current lack of ground truth datasets in this domain.

Finally, optimisation of telemetry collection and provenance verification mechanisms is necessary to reduce operational overhead and support deployment in resource-constrained environments.

# 6 Conclusions

This work introduced an execution-bound methodology for exploitability assessment in Agentic AI systems. By integrating AIBOM artefact modelling, runtime telemetry capture, and cryptographically sealed provenance into the CSAF-VEX lifecycle, the framework moves beyond static SBOM-based vulnerability enumeration. The results demonstrate that exploitability classification in AI-native environments depends on artefact activation paths, tool invocation surfaces, and enforced execution constraints rather than dependency presence alone.

The empirical evaluation showed measurable improvements in classification precision and false-positive reduction when model artefacts, prompt routing definitions, and tool connectors were explicitly bound to runtime evidence. In Agentic AI workloads, optional capabilities and unreachable invocation paths frequently distort static vulnerability assessments. The AIBOM extension enabled identification of such conditions, materially altering exploitability outcomes in AI-specific components while preserving recall.

Deterministic replay and provenance anchoring established advisory generation as a reproducible computational process. This property is particularly significant in AI deployments where artefact versions, model binaries, and orchestration logic may evolve independently. By binding model hashes, prompt commits, container digests, and policy states within a verifiable execution envelope, CSAF-VEX artefacts become execution-grounded attestations rather than declarative vendor statements.

The framework therefore provides a structured pathway for operationalising AIBOM within modern AI supply chains. It aligns artefact transparency, runtime validation, and advisory automation into a unified model suitable for containerised, multi-agent, and tool-using AI systems. Future research should extend this approach to incorporate semantic prompt-level threat modelling, training-time artefact assurance, and federated cross-organisational advisory verification.

**Dr. Petar Radanliev**
Parks Road,
Oxford OX1 3PJ
United Kingdom
Email: petar.radanliev@cs.ox.ac.uk
BA Hons., MSc., Ph.D. Post-Doctorate

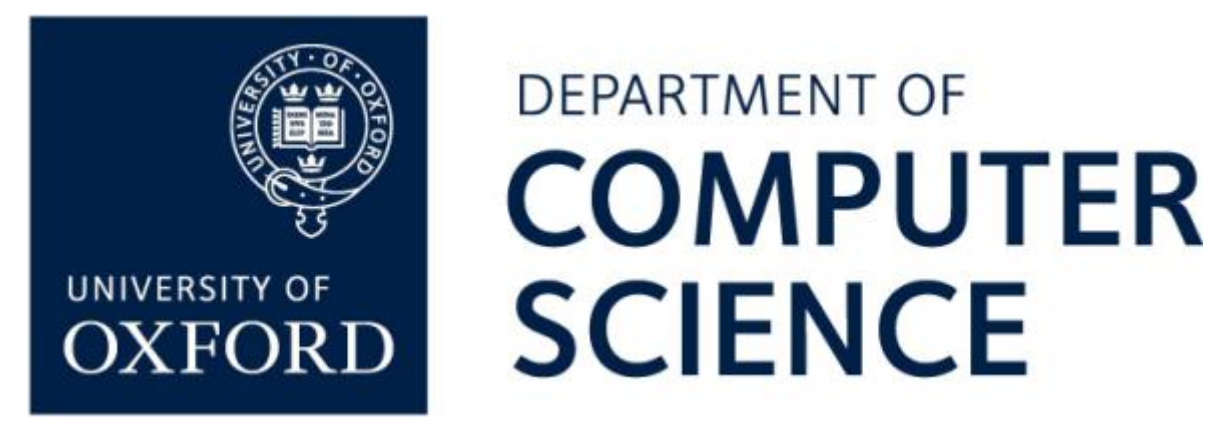

**Dr. Petar Radanliev**
Parks Road,
Oxford OX1 3PJ
United Kingdom
Email: petar.radanliev@cs.ox.ac.uk
BA Hons., MSc., Ph.D. Post-Doctorate

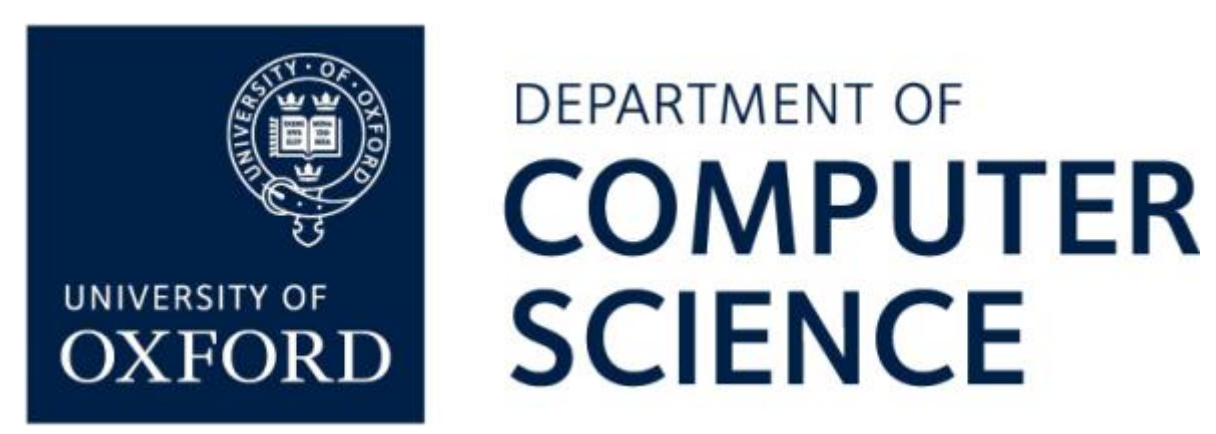

**Dr. Petar Radanliev**
Parks Road,
Oxford OX1 3PJ
United Kingdom
Email: petar.radanliev@cs.ox.ac.uk
BA Hons., MSc., Ph.D. Post-Doctorate

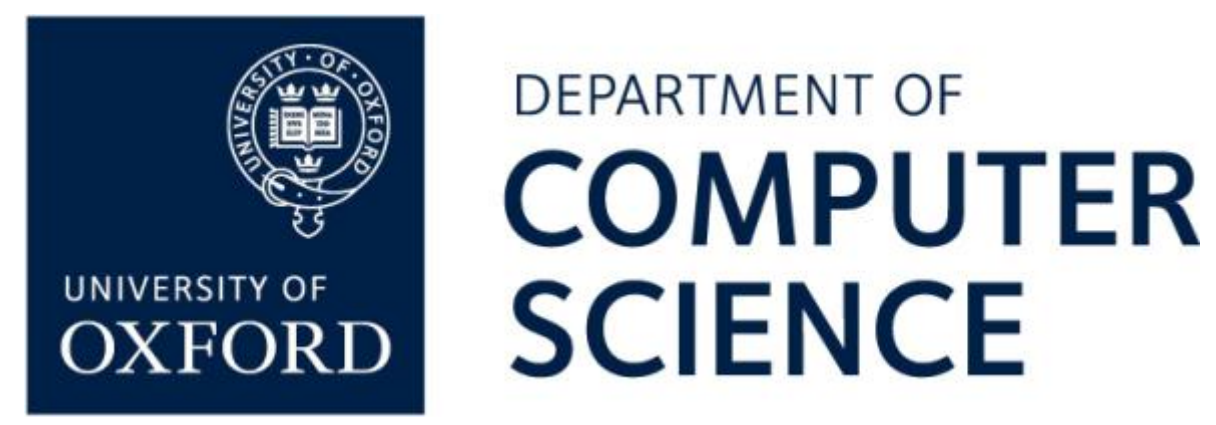

**Dr. Petar Radanliev**
Parks Road,
Oxford OX1 3PJ
United Kingdom
Email: petar.radanliev@cs.ox.ac.uk
BA Hons., MSc., Ph.D. Post-Doctorate

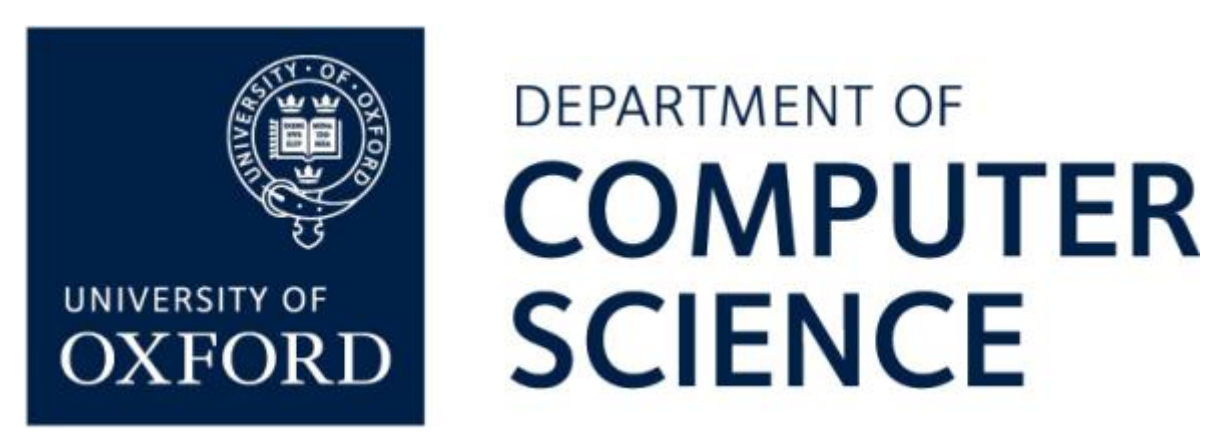

**Dr. Petar Radanliev**
Parks Road,
Oxford OX1 3PJ
United Kingdom
Email: petar.radanliev@cs.ox.ac.uk
BA Hons., MSc., Ph.D. Post-Doctorate

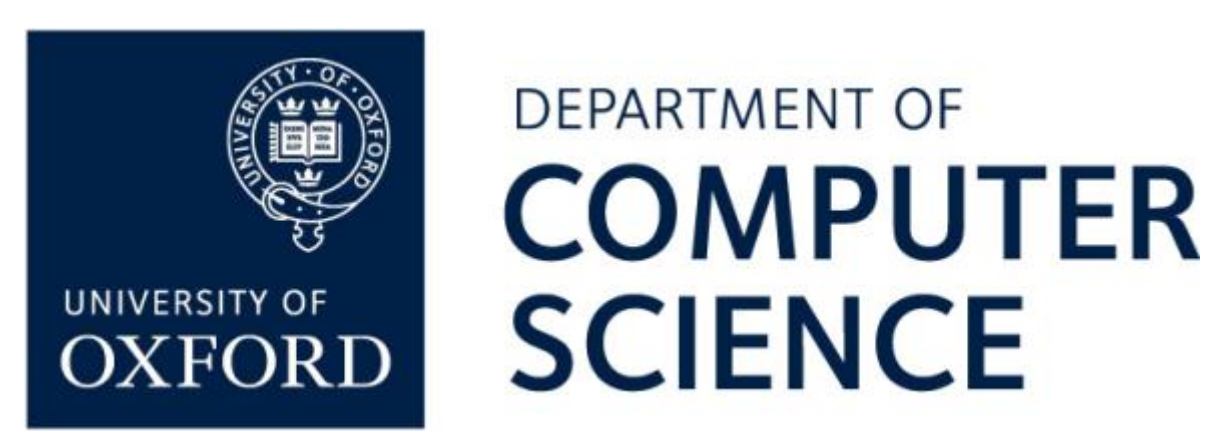

**Dr. Petar Radanliev**
Parks Road,
Oxford OX1 3PJ
United Kingdom
Email: petar.radanliev@cs.ox.ac.uk
BA Hons., MSc., Ph.D. Post-Doctorate

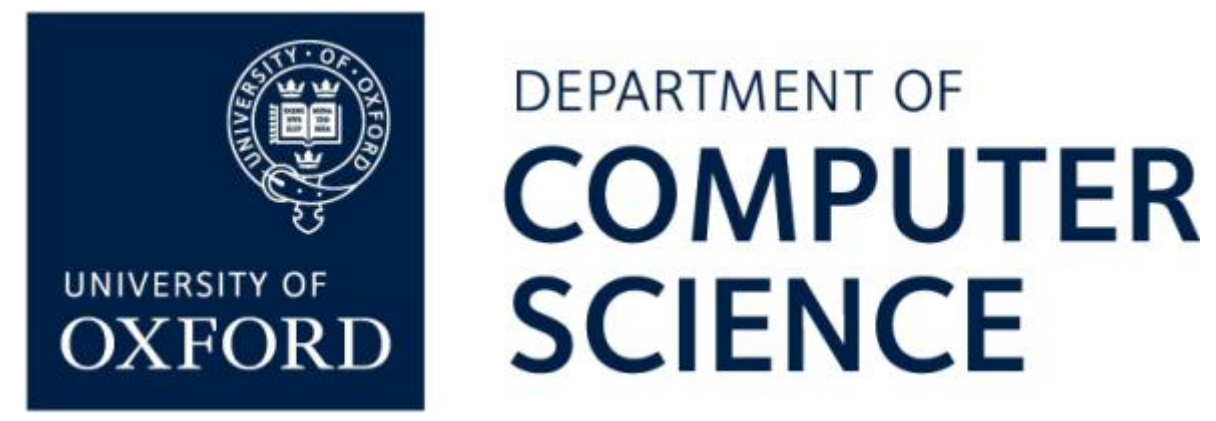

**Dr. Petar Radanliev**
Parks Road,
Oxford OX1 3PJ
United Kingdom
Email: petar.radanliev@cs.ox.ac.uk
BA Hons., MSc., Ph.D. Post-Doctorate

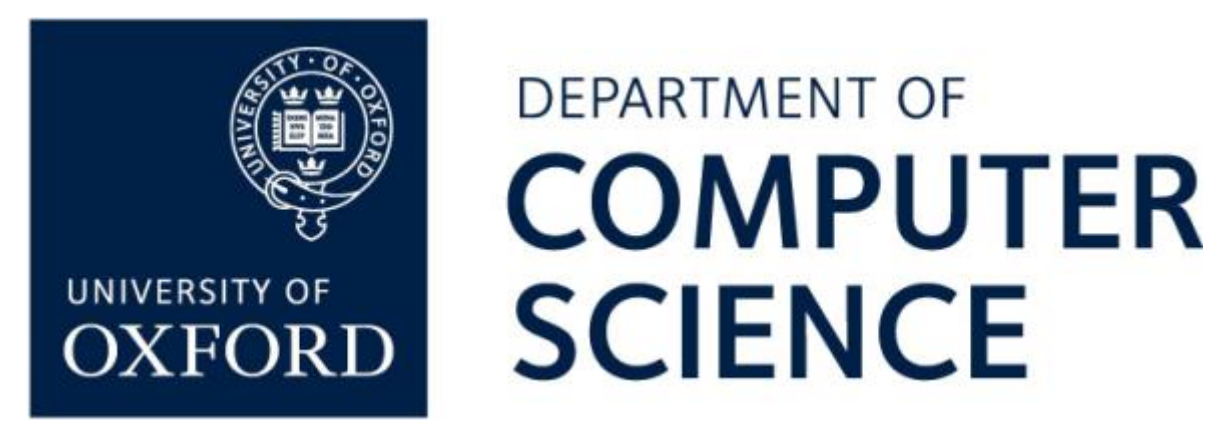